\documentclass[twocolumn,aps,pra,shownopacs,superscriptaddress]{revtex4}
\usepackage{graphicx}
\usepackage{amssymb}
\usepackage{amsmath}
\usepackage{amsthm}
\usepackage{newtxtext,newtxmath}
\usepackage{dsfont}
\usepackage{wasysym}
\usepackage{array}
\usepackage{multirow}
\usepackage{makecell}
\usepackage{xcolor}
\usepackage[colorlinks=true,linkcolor=blue,urlcolor=blue,citecolor=blue,pdfauthor={ },pdftitle={},pdfsubject={ },pdfkeywords={ }]{hyperref}
\graphicspath{{Figures/}}

\begin{document}
	
	\title{Optimal and robust experiment design for quantum state tomography of star-topology register}
	
	\date{\today}
\author{Ran Liu}
\affiliation{CAS Key Laboratory of Microscale Magnetic Resonance and School of Physical Sciences, University of Science and Technology of China, Hefei 230026, China}
\affiliation{CAS Center for Excellence in Quantum Information and Quantum Physics, University of Science and Technology of China, Hefei 230026, China}
\author{Yanjun Hou}
\affiliation{CAS Key Laboratory of Microscale Magnetic Resonance and School of Physical Sciences, University of Science and Technology of China, Hefei 230026, China}
\affiliation{CAS Center for Excellence in Quantum Information and Quantum Physics, University of Science and Technology of China, Hefei 230026, China}
\author{Ze Wu}
\affiliation{CAS Key Laboratory of Microscale Magnetic Resonance and School of Physical Sciences, University of Science and Technology of China, Hefei 230026, China}
\affiliation{CAS Center for Excellence in Quantum Information and Quantum Physics, University of Science and Technology of China, Hefei 230026, China}
\author{Hui Zhou}
\affiliation{School of Physics, Hefei University of Technology, Hefei, Anhui 230009, China}
\author{Jiahui Chen}
\affiliation{Department of Physics, University of Waterloo, Waterloo, Ontario N2L 3G1, Canada}
\affiliation{Institute for Quantum Computing, Waterloo, Ontario N2L 3G1, Canada}
\author{Xi Chen}
\affiliation{Department of Physics, The Hong Kong University of Science and Technology, Clear Water Bay, Kowloon, Hong Kong, China}
\author{Zhaokai Li}
\email{zkli@ustc.edu.cn}
\affiliation{CAS Key Laboratory of Microscale Magnetic Resonance and School of Physical Sciences, University of Science and Technology of China, Hefei 230026, China}
\affiliation{Hefei National Laboratory, Hefei 230088, China}
\affiliation{CAS Center for Excellence in Quantum Information and Quantum Physics, University of Science and Technology of China, Hefei 230026, China}
\author{Xinhua Peng}
\email{xhpeng@ustc.edu.cn}
\affiliation{CAS Key Laboratory of Microscale Magnetic Resonance and School of Physical Sciences, University of Science and Technology of China, Hefei 230026, China}
\affiliation{Hefei National Laboratory, Hefei 230088, China}
\affiliation{CAS Center for Excellence in Quantum Information and Quantum Physics, University of Science and Technology of China, Hefei 230026, China}
	
	\begin{abstract}
		\par While quantum state tomography plays a vital role in the verification and benchmarking of quantum systems, it is an intractable task if the controllability and measurement of quantum registers are constrained. In this paper, we study the quantum state tomography of star-topology registers, in which the individual addressability of peripheral spins is infeasible. Based on the star-symmetry, we decompose the Hilbert space to alleviate the complexity of tomography and design a compact strategy with minimum number of measurements. By optimizing the parameterized quantum circuit for information transfer, the robustness against measurement errors is also improved. Furthermore, we apply this method to a 10-spin star-topology register and demonstrate its ability to characterize large-scale systems. Our results can help future investigations of quantum systems with constrained ability of quantum control and measurement.

	\end{abstract}
	\maketitle

	\section{Introduction}
	 The estimation of an unknown quantum state, known as quantum state tomography (QST), is one of the fundamental problem in quantum science and technology \cite{BK1,BK2,BK3,BK4}. It has become an indispensable tool in validating and benchmarking quantum devices. Typically, standard QST requires measurements to be carried out in different settings to observe different parts of the density matrix. The switch between settings is usually accomplished by applying unitary operation before experimental measurements \cite{Litomo}.

     Most of the previous work consider the problem of QST based on quantum systems with ability of universal quantum control \cite{fullcontrl,fullcontrl2,fullcontrl3}. However, it is common in certain quantum systems that particles or qubits cannot be individually addressed, which poses restrictions to universal quantum operation and individual measurements. As an example, the Hong-Ou-Mandel effect \cite{Hongou} in the field of quantum optics leads to the result that the photons with the same characteristics enter the same mode and become indistinguishable. In spin systems, this also happens when the quantum spins cannot be addressed either by their positions or the magnet resonant frequencies, e.g., a large number of trapped ions \cite{trapped1,trapped2} and the nuclear spins with magnetic equivalence \cite{mageq1,mageq2}. 
    
    To tackle the problem of QST for such quantum system composed of indistinguishable particles, the permutationally invariant quantum tomography scheme has been proposed and developed \cite{PI,PItomo1,hidden,PItomo2,PItomo3,PItomo5}. Taking advantage of the permutationally invariant symmetry, the quantum states can be efficiently characterized and reconstructed, which is also scalable for multi-qubit systems.

	 \begin{figure*}[htb]
		\begin{center}
			\includegraphics[scale=0.58]{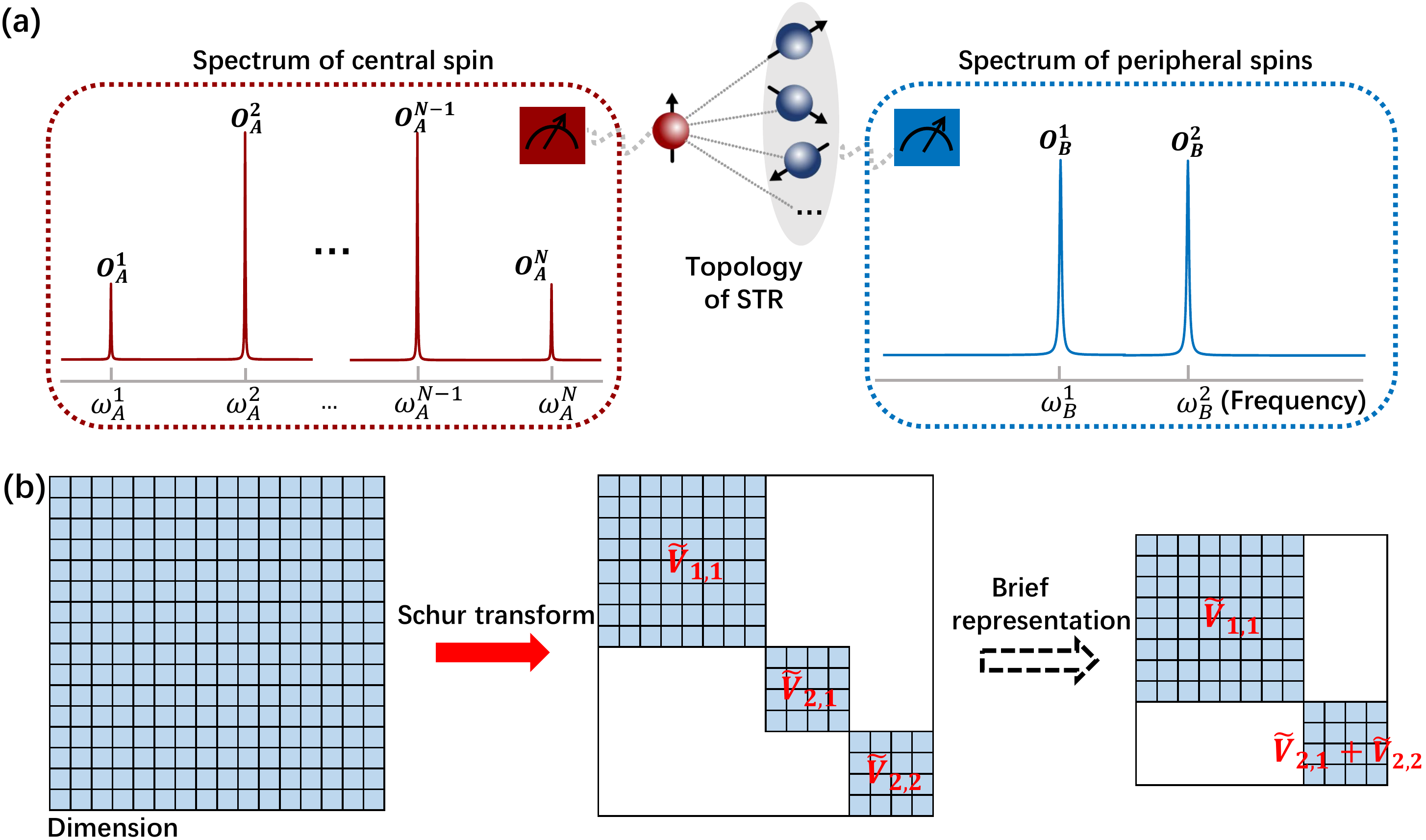} 
			\caption{(a) A schematic diagram of the topology of STR and the resonant spectra of spins polarized in $x/y$-axis. The frequencies and the corresponding observables are labeled. (b) Decomposed Hilbert space of 4-qubit STR by Schur transform. We further add the isomorphic subspaces together for a brief representation, for the information in isomorphic subspaces are the same.} \label{diagram}
		\end{center}
	\end{figure*}

    Star-topology register (STR) has a specific network topology that consists of a central spin uniformly interacting with a set of peripheral spins, and these peripheral spins cannot be individual addressed due to their magnetic equivalence. The indistinguishability of permutationally invariant particles means the ease of collectively manipulating large-scale quantum system. Besides, STR can be used to efficiently prepare large-scale entangled states, i.e. NOON state, due to its specific form of coupling. So STR has been widely used in quantum sensing and quantum simulation, such as measuring magnetic field \cite{apy1}, characterizing radio-frequency (RF) inhomogeneity \cite{apy2}, studying noise spectroscopy \cite{apy3}, thermodynamics of many-body systems \cite{apy4}, algorithmic cooling \cite{apy5} and temporal ordered phase \cite{apy6}. However, due to the specific star-symmetry, the controllability and measurement in STR are constrained \cite{sum}, and whether the STR quantum states can be fully determined by available measurement settings has not been investigated yet.

	In this work, we provide a novel scheme for quantum state tomography of STR. By exploiting the star-symmetry, we use an effective representation of quantum state in a decomposed Hilbert space, and design the optimal scheme with the minimal number of measurement settings. The parameterized quantum circuits (PQCs) used for information transfer are then optimized to improve the robustness against measurement errors. As a demonstration, we numerically simulate the QST for a 10-spin STR and show the feasibility and scalability of our method. The rest of the paper is structured as follows. In Sec.\ II we first give a brief introduction about STR, including its model, available measurements and decomposition of the Hilbert space. Then we describe our scheme of QST to be applied on STR in Sec.\ III. Sec.\ IV includes the error analysis of our method and the optimization strategy to improve the robustness. In Sec.\ V we move on to the numerical demonstration of our scheme in a 10-spin example and then conclude this work in Sec.\ VI.

	 \section{Star-topology register}
	 \subsection{Model}
	 An STR consists of a central spin (A) uniformly coupled to $N-1$ identical peripheral spins (M) with the same interaction strength $J_{AM}$, thus showing the so-called $star$-$symmetry$. A schematic diagram of the topology of STR is shown in Fig.\ \ref{diagram}(a). In the following, we consider the Ising-type interations between A and M with the strength of $J_{AM}$. The interactions among the peripheral spins are ineffective in our case due to magnetic equivalence symmetry, thus the system Hamiltonian can be written as

	\begin{eqnarray}
		H_0&=&H_A+H_M+H_{AM}\\\notag
		&=&\pi\omega_A\sigma_z^{A}+\pi\omega_M\sigma_z^{M}+\frac{\pi}{2} J_{AM}\sigma_z^{M}\sigma_z^{A}.
	\end{eqnarray}
Here $\omega_A$ and $\omega_M$ are the Larmor frequencies of central and peripheral spins caused by external magnetic field, which generally can be omitted in the rotating frame. $\sigma_\nu^{A}=\sigma_\nu^{A,N}$ is the Pauli matrix of central spin A and $\sigma_\nu^{M}=\sum_{i=1}^{N-1}\sigma_\nu^{M,i}$ is the collective Pauli matrix of peripheral spins M, where $\sigma_\nu^{\cdot,i}$ denotes the Pauli matrix of the $i$th spin with $\nu=\{x,y,z\}$. Here we label the central spin as the $N$-th (i.e., the last) qubit for the convenience of block diagonalization of the Hilbert space.

The additional control Hamiltonian arise from time-dependent magnetic field applied on $x$-$y$ plane. The central spin can be selectively addressed due to its distinct characteristics. Nevertheless, the peripheral spins are indistinguishable from one another and can only be collectively manipulated. Under the star-symmetry, the controllability of the system is not universal, and the control Hamiltonian can be written as
\begin{eqnarray}\label{contrlf}
	H_{\text{control}}&=&\pi u_x^A(t) \sigma_x^{A}+\pi u_y^A(t) \sigma_y^{A}\\\notag&+&\pi u_x^M(t) \sigma_x^{M}+\pi u_y^M(t)\sigma_y^{M},
\end{eqnarray}
where $\sqrt{u_x^{A/M}(t)^2+u_y^{A/M}(t)^2}$ is the instantaneous Rabi frequency determined by the strength of the control field.
 
\subsection{Measurement settings for STR}\label{spec}

The central spin in STR can be individually measured while the peripheral ones are indistinguishable. Without loss of generality, we consider the measurement setting for central spin being its polarization on $x$-$y$ plane as
\begin{eqnarray}
	 M_{A,x/y}=\sigma_{x/y}^{A}.
\end{eqnarray}
To extract more information about the quantum state, we can have the system evolve under the internal Hamiltonian $H_0$ and measure $M_{A,x/y}$ at different times. In this way, a sequential of time domain signals can be obtained as 
\begin{eqnarray}
\langle M_{A,x/y}\rangle(t)=\text{Tr}(e^{-iH_0t}\rho e^{iH_0t}\sigma_{x/y}^{A}),
\end{eqnarray}

where $\langle M_{A,x}(t)\rangle,\langle M_{A,y}(t)\rangle$ can both be extracted from the independent detection along $x$- and $y$-axis. After a Fourier transform on the sampling time domain signals, i.e.,
\begin{eqnarray}
S(\omega)=\int_0^\infty \langle M_{A,x/y}(t)\rangle(t) e^{-i\omega t}dt,
\end{eqnarray}
we can obtain the frequency spectrum $S(\omega)$, which shows $N$ peaks at the frequencies of $\omega_A^1=-\frac{N-1}{2}J_{AB},\omega_A^2=-\frac{N-3}{2}J_{AB},...,\omega_A^N=\frac{N-1}{2}J_{AB}$ . The corresponding observables for each peak can be written as

\begin{eqnarray}
O_A^i&=&\sum_l|n,N-1-n\rangle_l\langle n,N-1-n|_l\otimes \sigma_{x/y}^{A},
\end{eqnarray}
and equivalently the expectation of $O_A^i$ can be extracted from the signal on the corresponding peak. Here $|n,N-1-n\rangle$ is the quantum state of $N-1$ peripheral spins with $n$ spins being up and $N-1-n$ spins being down with $n=0,1,2...N-1$, and $l$ runs over the indistinguishable permutations of them. $S(\omega)$ obtained from state polarized in $x/y$-axis and the corresponding observables and frequencies are depicted in Fig.\ \ref{diagram}(a). 

Similarly, the collective observables for the peripheral qubits can be obtained as
\begin{eqnarray}
	 O_B^1&=&\sigma_{x/y}^{M}\otimes|0\rangle\langle0|,\\\notag
	 O_B^2&=&\sigma_{x/y}^{M}\otimes|1\rangle\langle1|.
\end{eqnarray}
The corresponding frequency spectrum is also depicted in Fig.\ \ref{diagram}(a). As we can see, the degenerate levels in $H_0$ lead to the specific N-peak spectrum of central spin and two-peak spectrum of peripheral ones, which also means that the individual detection of peripheral spins becomes infeasible. The set of observables 
\begin{eqnarray}
\{O_A^{1,2,...N},O_B^{1,2}\}
\end{eqnarray}
forms the initial measurement settings for STR.

\begin{figure*}[htb]
	\begin{center}
		\includegraphics[scale=0.6]{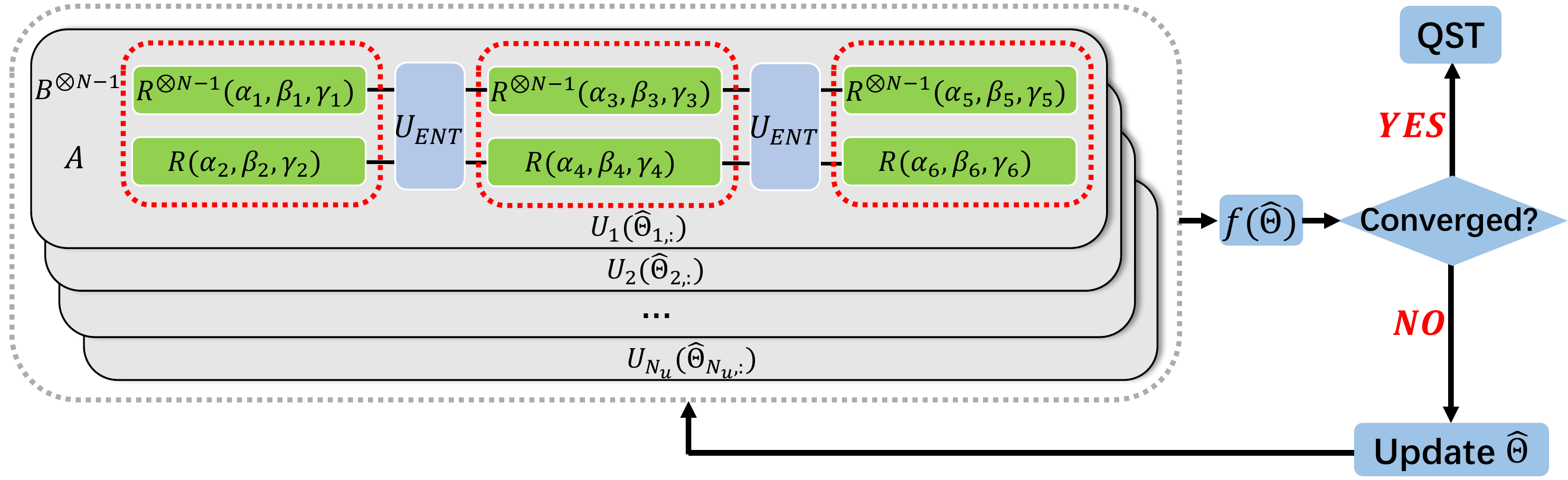} 
		\caption{Flowchart for optimizing readout operations for QST. Each of the gray boxes contains a PQC for generating readout operation with parameters to be optimized. Each row of $\hat\Theta$, i.e., $\hat\Theta_{1,2,3...N_u,:}$, corresponds to single-qubit rotation parameters (boxed in red dashed line) in one PQC. The cost function $f(\hat\Theta)$ can be calculated from $N_u$ readouts. 
	    To minimize $f(\hat \Theta)$, $\hat\Theta$ can be updated  until converged.}\label{opt}
	\end{center}
\end{figure*}
\subsection{Decomposition of the Hilbert space}\label{subs}

Due to the $N-1$ peripheral spins in STR can only be manipulated collectively, the available unitary controls are thus constrained. These unitary operators can be represented in the decomposed Hilbert space based on Lie algebra technique. The permutationally invariant Lie subalgebra of $\text{SU}(2^{N-1})$ can be defined as
\begin{eqnarray}
	\mathcal L_{\text{PI}}=\{a|a\in\text{SU}(2^{N-1}),\text{such that }\forall\Pi\in S_{N-1},\Pi a\Pi=a\}.
\end{eqnarray}
Here $S_{N-1}$ is the symmetric group of all permutations of the $N-1$ peripheral spins, and $\Pi$ is one of its elements. For the Hilbert space of peripheral spins written as $(\mathbb{C}^2)^{\otimes N-1}$, $\mathcal L_{\text{PI}}$ can be fully reduced under the following decomposition
\begin{eqnarray}
	(\mathbb{C}^2)^{\otimes N-1}\cong \bigoplus_{i=1}^{\lceil N/2\rceil}\bigoplus_{j=1}^{N_i} V_{i,j}.
\end{eqnarray}
Here 
\begin{eqnarray}
	N_i=\frac{N+2-2i}{N+1-i}C_{N-1}^{i-1},
\end{eqnarray}
with $C_{N-1}^{i-1}$ as the binomial number. Each $V_{i,j}$ is an irreducible subspace of $(\mathbb{C}^2)^{\otimes N-1}$ for $\mathcal L_{\text{PI}}$ with the dimensions of $2i$ when $N$ is even and $2i-1$ when $N$ is odd, and $N_i$ is the number of isomorphic irreducible subspaces \cite{jiahui}. The largest subspace is known as Dicke subspace, which associated with total angular
momentum $(N-1)/2$.

The basis transformation of the decomposition, defined as $U_{\text{tr}}$, is known as the Schur transform \cite{Schur}. Similarly, the evolved quantum states with star-symmetry possess the same symmetry and can be represented in the same decomposed Hilbert space. This block-decomposition represents a natural way to treat permutationally invariant states \cite{PItomo1,hidden,PItomo2,PItomo3}.

Considering the whole Hilbert space of STR, it can then be block diagonalized by applying $U_{\text{tr}}\otimes\mathds{1}_2$, where the left distribution law over direct sum \cite{distribu}, i.e., 
\begin{eqnarray}\label{divspc}
(\bigoplus_{i,j}V_{i,j})\bigotimes V_C=\bigoplus_{i,j}(V_{i,j}\bigotimes V_C)\equiv\bigoplus_{i,j}\widetilde{V}_{i,j}
\end{eqnarray}
is used. Correspondingly, the quantum states can be written as
\begin{eqnarray}
\rho=\bigoplus_{i,j}\varrho_{i,j}=\bigoplus_{i,j}\lambda_{i,j}\rho_{i,j},
\end{eqnarray}
where $\lambda_{i,j}=\text{Tr}(\varrho_{i,j})$. Due to $\widetilde{V}_{i,j}$ and $\widetilde{V}_{i,j'}$ are isomorphic and contain the same information of quantum state, so we have $\lambda_{i,1}= \lambda_{i,2}=...\equiv \lambda_{i,:}$.

A schematic diagram of the decomposed Hilbert space of 4-qubit STR is shown in Fig.\ \ref{diagram}(b). We also give a brief representation $\rho_\text{br}$ for such quantum states by adding the isomorphic subspaces together, as it is convenient for exhibiting quantum states when $N$ is large.
 
\section{Quantum State Tomography of STR}   

A QST problem can be converted into a multiparameter estimation one. Let $\{\mathcal B_m\}$ denote a set of Hermitian operators satisfying $\mathrm{Tr}(\mathcal B_m\mathcal B_n)=0$ when $m\neq n$. We suppose that the quantum state to be constructed can be completely represented by $\{\mathcal B_m\}$:
\begin{eqnarray}\label{state}
	\rho=\frac{\mathds{1}}{2^N}+\sum_m c_m\mathcal B_m,
\end{eqnarray}
where $\mathds{1}$ is the identity matrix and $c_m=\mathrm{Tr}(\rho\mathcal B_m)/\mathrm{Tr}(\mathcal B_m\mathcal B_m)$. For the general quantum states, due to the constrains that $\rho$ is Hermitian with unit trace, the number of degrees of freedom is $4^N-1$.

In a tomographic experiment, multiple copies of the state to be constructed are prepared and measured under a specific measurement setting, which corresponds to a set of experimental observables $\{  O_i\}_{i=1}^{N_o}$. Typically, the data collected from a single setting is generally not informationally complete to reconstruct $\rho$, hence the results from multiple settings are needed. To implement the switch between different settings, a set of unitary readout operations $\{U_j\}_{j=1}^{N_u}$ can be applied before measurements; that is, $\text{Tr}(U_j\rho U_{j}^\dagger O_i)=\text{Tr}(\rho U_j^\dagger  O_iU_j)$ \cite{NMRtomo}. So the available sets of observables can be noted as $\{\tilde{O}_k=U_j^\dagger O_iU_j\}_{i=1;j=1}^{N_o;N_u}$. The measurement result $o_k$ of the $k$th measurement $\tilde{O}_k$ is
\begin{eqnarray}\label{timecon}
	o_k=\text{Tr}(\rho\tilde{O}_k)=\sum_mc_m\text{Tr}(U_j^\dagger O_iU_j\mathcal B_m),
\end{eqnarray}
and a system of linear equations can be obtained,
\begin{eqnarray}\label{prop}
	\vec o=\hat{\mathcal F}\cdot\vec c.
\end{eqnarray}
Here $\vec o=(o_1,o_2,...o_k,...)^{\rm T}$, and $\hat{\mathcal F}$ is the transfer matrix with the element $\hat{\mathcal F}_{km}=\text{Tr}(\tilde{O}_k\mathcal B_m)$. By solving Eq. \eqref{prop},
the solution, i.e., $\vec c=(c_1,c_2,...c_m,...)^{\rm T}$, is unique (or the state can be fully determined) if and only if the rank of $\hat{\mathcal F}$ equals to the dimension of $\vec c$. So in the general case, the minimum of $N_u$ to fully determine $\rho$ is $\lceil\frac{4^N-1}{N_o}\rceil$.

We then move to the case of STR. Due to the star-symmetry, the exponential scaling of unknown parameters can be effectively alleviated. In the following, we only consider the case that the evolved quantum states can always be represented in $\lceil N/2\rceil$ different irreducible subspaces as mentioned in Sec. \ref{subs} and the trace is preserved in each subspace. The simplest case is the unitary evolution under Hamiltonian with star-symmetry. So the number of degrees of freedom for STR quantum states is
\begin{eqnarray}
\begin{split}
	\left \{
	\begin{array}{ll}
		\sum_{i=1}^{N/2}(4i)^2-\frac{N}{2},      &\text{when N is even}\\
		\sum_{i=1}^{(N+1)/2}(4i-2)^2-\frac{N+1}{2},     &\text{when N is odd}
	\end{array}\right.
\end{split}
\end{eqnarray}
The result for both cases can be expressed as
\begin{eqnarray}\label{Nc}
\frac{2N(N+1)(N+2)}{3}-\lceil\frac{N}{2}\rceil,
\end{eqnarray}
where $\lceil\frac{N}{2}\rceil$ comes from the trace-preserving conditions in different subspaces, i.e.,
\begin{eqnarray}\label{trinf}
\left\{\text{Tr}(\varrho_{i,:})=\lambda_{i,:}|i=1,2,...\lceil N/2\rceil\right\}.
\end{eqnarray}
Due to the identity matrix in each subspace commutes with readout operations $U$ and observables $O$, the trace in each subspace thus have no effect on the measurement results. Here we take the trace in each subspace, as denoted in Eq. \eqref{trinf}, as a prior information, so another $\lceil N/2\rceil$ equations can be obtained and absorbed into Eq. \eqref{prop}.

According to Sec. \ref{spec}, the number of elements in $\{O_i\}$ (i.e., $N_o$) for a single measurement setting is $2N+4$. As a result, the minimum of $N_u$ for quantum state tomography of STR is
\begin{eqnarray}\label{minN}
	N_{u}^{\text{min}}=\left\lceil\frac{\frac{2N(N+1)(N+2)}{3}-\lceil\frac{N}{2}\rceil}{2N+4}\right\rceil\sim N^2.
\end{eqnarray}

Typically, QST requires a set of readout operations $\{U_j\}_{j=1}^{N_u}$ to be selected with $N_u$ as small as possible and $\hat{\mathcal F}$ is column full rank. Previous protocols typically considered it as a set cover problem——$\rho$ can be fully determined iff $\{\tilde{O}_k\}$ can cover $\{\mathcal B_i\}$. To solve this NP-hard problem, greedy strategy \cite{greed} and integer programming \cite{Litomo} have been investigated. Both of them are promising to fully reconstruct $\rho$ with the minimum number of $U$, while how the choice of $U$ influences the precision of tomography has not been considered. In the following section, we propose a novel scheme for QST, in which we use the minimal readout operations generated by parameterized quantum circuits (PQCs) and then optimize them to improve the robustness against measurement errors.

\section{Improve the robustness of QST by optimizing readout operations}

Intuitively, readout operations for QST are supposed to transfer tomographically complete information into available measurement settings, and they can be achieved by suitably chosen random PQCs \cite{PQC0}. What's more, PQCs are flexible to be further optimized for specific demands by tuning the parameters. Here, we generate informationally complete readout operations by random PQCs and then optimize them to improve the robustness against measurement errors.

The structure of PQCs is illustrated in Fig.\ \ref{opt}, where two types of quantum operations, i.e., single-qubit rotations and entangling gates, are included. A single-qubit rotation can be determined by 3 independent parameters, which is denoted as $R(\alpha,\beta,\gamma)\equiv R_x(\alpha)R_y(\beta)R_x(\gamma)$ with $R_\nu(x)=\text{exp}(-ix\sigma_\nu/2)$. Due to the peripheral spins can only be manipulated collectively, a layer of single-qubit rotation involves 6 rotation parameters as boxed by the red dashed lines. An entangling layer is realized by the free evolution under the system Hamiltonian with the interval $\tau=1/2J_{AM}$. Here we denote a $n-$layer PQC as a one with the first layer being a single-qubit rotation while the others consisting of an entangling gate and a single-qubit rotation. Enough PQCs with sufficiently long sequence of these layers can produce the column full rank transfer matrix $\hat{\mathcal F}$. When the structure of PQCs is specified, $\hat{\mathcal F}$ can be determined by the parameter matrix $\hat\Theta$ of PQCs, where each row of $\hat\Theta$, i.e., $\hat\Theta_{1,2,3...N_u,:}$, corresponds to parameters in one PQC.

With such transfer matrix, the quantum state can be reconstructed by calculating the vector of variables $\vec c$,
\begin{eqnarray}\label{solv}
	\vec c=\hat{\mathcal F}^{-1}\cdot\vec o. 
\end{eqnarray}
This reconstruction is computationally simple and known as linear inversion \cite{LI}. When the matrix $\hat{\mathcal F}$ is not a square one, or it's singular, $\hat{\mathcal F}^{-1}$ can be replaced by Moore-Penrose pseudoinverse \cite{Moore}. For a variable, e.g., $c_m$, in $\vec c$, we have
\begin{eqnarray}\label{solvs}
	c_m=\sum_{k=1}^{N_oN_u+\lceil N/2\rceil}(\hat{\mathcal F}^{-1})_{mk}o_k,
\end{eqnarray}
from which the error of experimental measurement on $o_k$ is transferred into $c_m$. According to the error propagation formula, we have
\begin{eqnarray}
	\text{Var}(c_m)&=&\sum_{k=1}^{N_oN_u+\lceil N/2\rceil}\left|\frac{\partial c_m}{\partial o_k}\right|^2\times\text{Var}(o_k)\notag\\
&=&\sum_{k=1}^{N_oN_u+\lceil N/2\rceil}|\mathcal F^{-1}_{mk}|^2\times \text{Var}(o_k),
\end{eqnarray}
where $\text{Var}(c_m)$ and $\text{Var}(o_k)$ are the variances of $c_m$ and $o_k$, respectively.

To minimise the variance of $c_m$ and improve the precision of QST, the parameters in PQCs for generating readout operations can be optimized. The cost function can be defined as the weighted sum of $\text{Var}(c_m)$, i.e.
\begin{eqnarray}\label{costfun}
f(\hat\Theta)&=&\sum_mw_m\text{Var}(c_m)\\\notag
&=&\sum_mw_m\sum_{k=1}^{N_oN_u+\lceil N/2\rceil}|[\hat{\mathcal F}^{-1}(\hat \Theta)]_{mk}|^2\times \text{Var}(o_k).
\end{eqnarray}
Here $w_m$ is the weight factor and can be defined according to specific demand. Combining with specific optimizing strategy, $\hat \Theta$ can be updated to minimise $f(\hat\Theta)$ until converged as shown in Fig.\ \ref{opt}.

\begin{figure*}[htb]
	\begin{center}
		\includegraphics[scale=0.65]{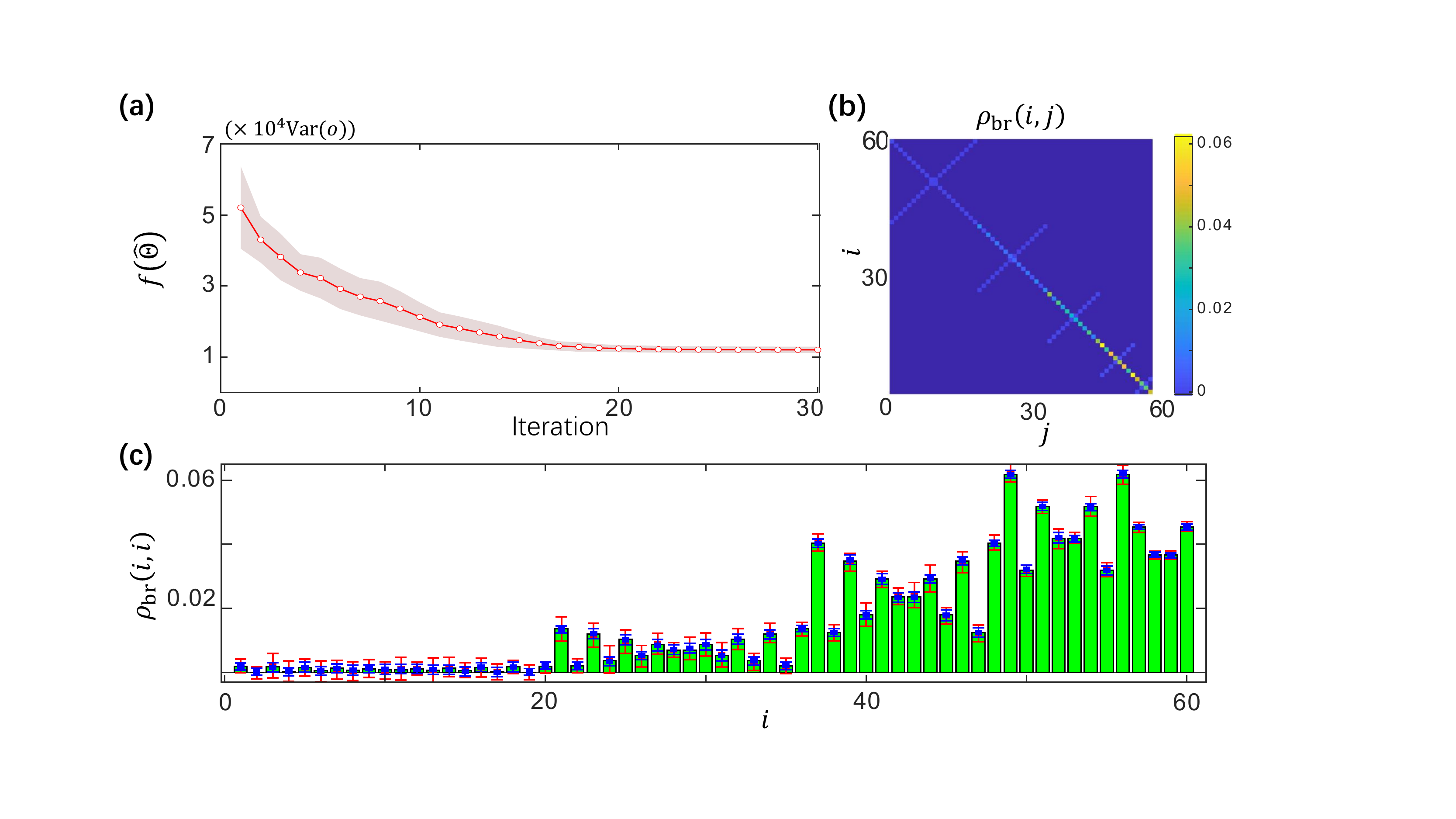} 
		\caption{(a) The process of minimizing the cost function $f(\hat\Theta)$ to improve the robustness against measurement errors. Statistical result over 10 runs of numerical optimization. The red solid line is the average while the dashed area is the standard deviation. (b) Real part of the brief representation of the 10-qubit density matrix $\rho$ to be determined. (c) Diagonal elements of reconstructed density matrix $\rho$. The green bars are the ideal values of $\rho_\text{br}(i,i)$, while the error bars contain the average and standard deviation of simulated reconstruction over 100 repetitions. Here the red and blue error bars correspond to results before and after the optimization of PQCs, respectively}\label{exam}
	\end{center}
\end{figure*}

\section{Example of 10-qubit STR}
The example of 10-qubit STR in this work is trimethyl phosphite (TMP) molecule consisting of a single $^{31}\text{P}$ nucleus and nine identical $^1\text{H}$ nuclei. Due to its large spin-cluster and clear spectra in NMR experiment, 10-qubit STR has been widely applied into the investigation of quantum sensing \cite{apy1}, measuring translation diffusion constant, mapping RF inhomogeneity \cite{apy2}, noise spectroscopy \cite{apy3} and thermodynamics of many-body systems \cite{apy4}. In the following, we take the 10-qubit STR as an example and give the detailed procedures for implementing our QST scheme.

\begin{itemize}
    \item[1.]\textbf{Choice of basis for STR quantum states.} 
    
    According to the decomposition of Hilbert space mentioned in Eq. \eqref{divspc}, we choose the set of basis for subspaces when $i=p$ as $\bigoplus_{j=1}^{N_p}B^p_{q_1,q_2}$, where $B^p_{q_1,q_2}$ is
	\begin{eqnarray}
    \begin{split}
	\left \{
	\begin{array}{ll}
		|q_1\rangle^p\langle q_1|^p, &1\le q_1=q_2\le \text{dim}(B^p)\\
		|q_1\rangle^p\langle q_2|^p+H.c.,     &1\le q_1<q_2\le \text{dim}(B^p)\\
		\text{i}|q_1\rangle^p\langle q_2|^p+H.c.,     &1\le q_1<q_2\le \text{dim}(B^p)\\
	\end{array}\right.
    \end{split}
    \end{eqnarray}
    Here $i=1,2,...\lceil N/2\rceil$ labels the irreducible subspace with different dimensions, $j=1,2,...N_i$ labels the isomorphic subspace, $|q_1\rangle^p,|q_2\rangle^p$ are the computational bases of the $p$-th subspace and $\text{dim}(B^p)$ denotes the dimension of $B^p$. Further, we have the set of basis for the whole Hilbert space as
    \begin{eqnarray}
		\mathcal B_m&=&\mathcal B^{p}_{q_1,q_2}\notag\\&\equiv&\frac{\delta_{ip}}{N_p}\bigoplus^{\lceil N/2\rceil}_{i=1}\bigoplus_{j=1}^{N_i}B^p_{q_1,q_2},
	\end{eqnarray}
	 where $\delta_{ip}$ denotes the Kronecker delta symbol. Apart from utilizing the decomposed Hilbert space, we also avoid introducing extra degrees of freedom caused by the repetitive information in isomorphic spaces under this basis, thus providing a compact representation of STR quantum states.

    \item[2.]\textbf{Generation of readout operations.} For 10-qubit STR, the number of degrees of freedom can be obtained from Eq. \eqref{Nc} as 875. Consequently, according to Eq. \eqref{minN} we have the minimal number of readout operations required to fully determine the state is 37. These readout operations generated by 3-layer PQCs, as shown in Fig.\ \ref{opt}, are enough to generate full rank transfer matrix $\hat{\mathcal F}(\hat\Theta)$.

	\item[3.]\textbf{Optimization of PQCs.} The robustness against noise can be quantified by the cost function $f(\hat\Theta)$ as mentioned in Eq. \eqref{costfun}. Here the variances of different experimental measurements $\text{Var}(o_k)$ are all taken the same value as $\text{Var}(o)$, and the weighted factors $w_m$ are all set as 1 for simplicity. $f(\hat\Theta)$ can then be solely determined by $\hat{\mathcal F}(\hat\Theta)$. The sequential quadratic programming method is adopted for optimization. At each iteration, the search direction is the solution of a quadratic programming subproblem \cite{SQP}, and the iteration is stopped when its number reaches 30. The optimization is repeated for 10 times with randomly initialized PQCs and the statistical results over the 10 runs are depicted in Fig.\ \ref{exam}(b), where the average of $f(\hat \Theta)$ decreases from $5.21\times10^4\text{Var}(o)$ to $1.20\times10^4\text{Var}(o)$. So the total variance transferred to $\vec c$ is significantly reduced after optimization. The optimization procedure is independent of the form of quantum state, so the PQCs can be tuned appropriately ahead of experimental measurements and then directly applied to the tomography of different quantum states. 
	
	\begin{table*}\caption{Precision improvement of tomography for different quantum states before and after the optimization of PQCs. These states have been widely investigated in quantum information and quantum metrology \cite{mixedstt,purestt}.}
\label{tab1}
\begin{tabular}{|cc|c|cc|cc|cc|}
\hline
\multicolumn{2}{|c|}{\multirow{2}{*}{Quantum State}}    & \multirow{2}{*}{\makecell[c]{Number of\\readout}} & \multicolumn{2}{c|}{$f(\hat\Theta)/\text{Var}(o_k)$}                                      & \multicolumn{2}{c|}{Infidelity (\%)}    & \multicolumn{2}{c|}{Distance ($\times10^{-2}$)}    \\ \cline{4-9} 
\multicolumn{2}{|c|}{}                     &                  &  \multicolumn{1}{c|}{Random}                  &     Optimized              & \multicolumn{1}{c|}{Random} & Optimized  & \multicolumn{1}{c|}{Random} & Optimized \\ \hline
\multicolumn{1}{|c|}{\multirow{4}{*}{Normal STR states}} & $S$ & \multirow{4}{*}{37} & \multicolumn{1}{c|}{\multirow{4}{*}{$5.11\times10^4$}} & \multirow{4}{*}{$1.08\times10^4$} & \multicolumn{1}{c|}{2.5$\pm$0.7} & 1.5$\pm$0.3 & \multicolumn{1}{c|}{6.4$\pm$1} & 3$\pm$0.3 \\ \cline{2-2} \cline{6-9} 
\multicolumn{1}{|c|}{} & $Cl$ &                   & \multicolumn{1}{c|}{}                  &                   & \multicolumn{1}{c|}{2.6$\pm$0.7} & 1.6$\pm$0.3 & \multicolumn{1}{c|}{6.4$\pm$1} &  3$\pm$0.3\\ \cline{2-2} \cline{6-9} 
\multicolumn{1}{|c|}{}                  & $Q1$ &                   & \multicolumn{1}{c|}{}                  &                   & \multicolumn{1}{c|}{2.4$\pm$0.7} & 1.6$\pm$0.3 & \multicolumn{1}{c|}{6.3$\pm$1} & 2.9$\pm$0.3 \\ \cline{2-2} \cline{6-9} 
\multicolumn{1}{|c|}{}                  & $Q2$  &                   & \multicolumn{1}{c|}{}                  &                   & \multicolumn{1}{c|}{2.6$\pm$0.8} & 1.6$\pm$0.3 & \multicolumn{1}{c|}{6.3$\pm$1} & 3$\pm$0.3 \\ \cline{1-9} 
\multicolumn{1}{|c|}{\multirow{3}{*}{States in the Dicke subspace}}
  
& Coherent&  \multirow{3}{*}{17}                 & \multicolumn{1}{c|}{\multirow{3}{*}{1.7$\times10^4$}} & \multirow{3}{*}{5.0$\times10^3$} & \multicolumn{1}{c|}{1.2$\pm$0.4} & 0.73$\pm$0.1 & \multicolumn{1}{c|}{2.2$\pm$0.6} & 1.3$\pm$0.2 \\ \cline{2-2} \cline{6-9} 
\multicolumn{1}{|c|}{}                  & GHZ &                   & \multicolumn{1}{c|}{}                  &                   & \multicolumn{1}{c|}{1.2$\pm$0.2} & 0.71$\pm$0.2 & \multicolumn{1}{c|}{2.1$\pm$0.4} & 1.2$\pm$0.2 \\ \cline{2-2} \cline{6-9} 
\multicolumn{1}{|c|}{}                  & Squeezed&                   & \multicolumn{1}{c|}{}                  &                   & \multicolumn{1}{c|}{1.3$\pm$0.3} & 0.71$\pm$0.1 & \multicolumn{1}{c|}{2.3$\pm$0.5} & 1.2$\pm$0.2 \\ \hline
\end{tabular}
\end{table*}
	\item[4.]\textbf{Reconstruction of the quantum state.} 
	Here we give a demonstration with the quantum state to be constructed $\rho$ as a mixture of 'many, some+some, many,' or MSSM states $\rho_\text{MSSM}$ in \cite{apy1},
	\begin{eqnarray}
	\rho_\text{MSSM}=&\sum_l&(|M,S\rangle_l|0\rangle+|S,M\rangle_l|1\rangle)\notag\\&\otimes&(\langle M,S|_l\langle0|+\langle S,M|_l\langle1|).
	\end{eqnarray}
	This state was used to measure magnetic field and can beat the standard quantum limit. In Fig.\ \ref{exam}(b), we exhibit the $2^{10}$-dimensional density matrix by using the brief representation $\rho_\text{br}$ mentioned in \ref{subs}, where the real part of $\rho_\text{br}$ is depicted. By applying the transfer matrix to experimental measurements as given by Eq. \eqref{solv} The quantum state can be reconstructed.
    \item[5.]\textbf{Error analysis.} 
	Imperfect experimental measurements can lead to deviations from the ideal results. We numerically simulate the deviations by introducing artificial fluctuations on each $o_k$ with the standard deviation as $\text{Var}(o)=3\times10^{-4}$. Due to the ranges of the amplitude of each observable $o_k$ are different, we define the relative standard deviation for each observable as
	\begin{eqnarray}
	\text{rsd}(o_k)\equiv\frac{\overline{\text{sd}(o_k)}}{\overline{|o_k|}},
	\end{eqnarray}
	where $\text{sd}(\cdot)$ is the standard deviation of the $k$th observable and the average is over $N_u$ different readout operations, respectively. Here, the relative standard deviations corresponding to the 12 observables of the 10-spin STR are 
	\begin{itemize}
	    \item[$\bullet$] central spin:\\
	    18.0\%, 22.1\%, 22.1\%, 16.4\%, 1.1\%\\
	    1.4\%, 17.0\%, 21.5\%, 22.1\%, 16.8\%\\ 
  
	    \item[$\bullet$] peripheral spins:\\
	    1.1\%, 1.2\%.
	\end{itemize}
According to Eq. \eqref{solv}, unknown quantum state can be solved from the simulated measurements. We repeat this process 100 times and calculate the average and standard deviation of estimated $c_m$, and the result of diagonal elements is shown in Fig.\ \ref{exam}(c). The green bars are the ideal values of $c_m$, while the error bars contain the average and standard deviation of simulated reconstruction. Here the red and blue error bars correspond to results before and after the optimization of PQCs, respectively.
To further quantify the precision, we calculate the infidelity and distance between reconstructed states and the ideal one. The average of infidelity over the 100 repetitions decrease from $2.4\%$ to $1.6\%$ after the optimization. Here the infidelity between the mixed states $\rho_1$ and $\rho_2$ is defined as
\begin{eqnarray}
1-F(\rho_1,\rho_2)\equiv1-\left(Tr\sqrt{\sqrt{\rho_1}\rho_2\sqrt{\rho_1}}\right)^2.
\end{eqnarray}
We also use the distance obtained from the Frobenius norm to quantify the improvement, which is defined as
\begin{eqnarray}
D(\rho_1,\rho_2)\equiv\sqrt{\text{Tr}[(\rho_1-\rho_2)(\rho_1-\rho_2)^\mathrm{T}]}.
\end{eqnarray}
The distance over the 100 repetitions are 0.063 and 0.029 before and after the optimization, respectively.

\end{itemize}

To further demonstrate the broad applicability of our scheme, we also apply it to some other typical 10-qubit STR states that have been widely investigated in quantum information and quantum metrology. For example, the infidelity of mixed states generated by standard strategy ($S$), classical strategy ($Cl$), quantum strategy 1 ($Q1$) and quantum strategy 2 ($Q2$) in \cite{mixedstt} decrease from around $2.6\%$ to $1.6\%$ after optimization, and the distance decrease from around $0.064$ to $0.030$. Here, the state generated with ($Q1$) is the same as the one in \cite{apy1}. 

Apart from improving the robustness against noise, our method is flexible to be adjusted to minimize the experimental effort by combining with prior information of unknown quantum states. For states in the Dicke subspace, the number of degrees of freedom becomes $(4N^2-1)=399$. As a consequence, the minimal number of readout operations needed can be reduced to $\lceil (4N^2-1)/(2N+4)\rceil=17$ when $N=10$. By optimizing these operations initially generated by random PQCs, $f(\hat\Theta)$ is decreased from $1.7\times10^4\text{Var}(o)$ to $5.0\times10^3\text{Var}(o)$. We give the demonstration with coherent spin state, GHZ state and spin-squeezed state in \cite{purestt}. 
Under the same standard deviation on $o_k$, 
the average of infidelity over the 100 repetitions decrease from $1.2\%$ to $0.72\%$ after the optimization, while the distance decreases from 0.072 to 0.012. The detailed information is shown in Table \ref{tab1}.

\section{Other strategies for improving the precision of tomography}
Apart from optimizing the parameters of PQCs to improve the robustness against the measurement errors, we can also change the structure or the number of PQCs for improving the precision of tomography.

\begin{figure*}[htb]
	\begin{center}
		\includegraphics[scale=0.54]{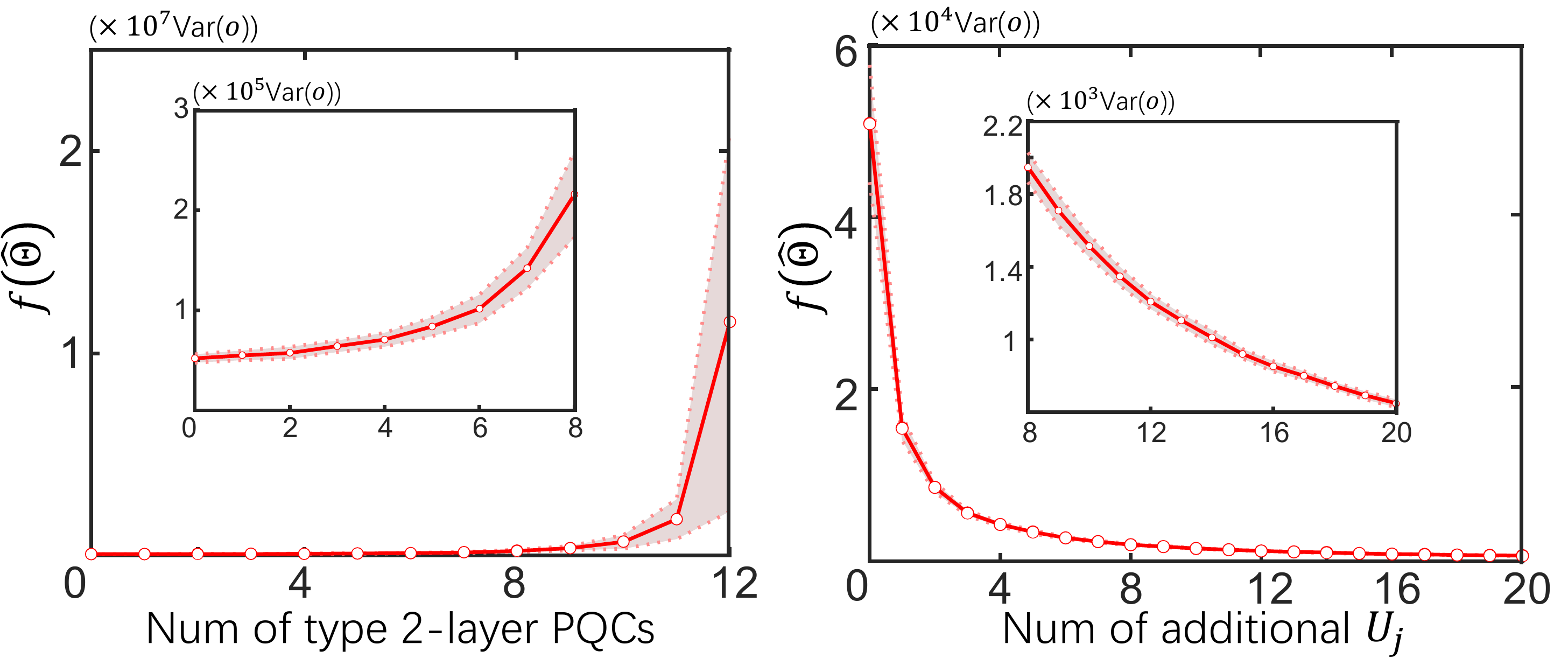} 
		\caption{The average and standard deviation of cost function $f(\hat\Theta)$ versus (a) the number of 2-layer PQCs in 37 readout operations and (b) additional readout operations apart from the 37 ones. The statistical results is based on 100 sets of readout operations generated by random PQCs.}\label{othStr}
	\end{center}
\end{figure*}
Specifically, in the example of 10-spin STR, 37 readout operations generated by 3-layer PQCs are employed to produce the column full rank transfer matrix $\hat{\mathcal{F}}$. While if part of them are replaced by PQCs with simpler structure, i.e., 2-layer or 1-layer PQCs, $\hat{\mathcal{F}}$ can still be column full rank. Here we only consider the combination of 2-layer and 3-layer PQCs. When the number of 2-layer PQCs is no more than 12, $\hat{\mathcal{F}}$ can be column full rank. The statistical results of the cost function $f(\hat\Theta)$ under different combination of these two types are shown in Fig.\ \ref{othStr}(a). For each combination, the average and the standard deviation over 100 sets of rand PQCs is depicted. $f(\hat\Theta)$ shows a growing tendency as the structure of PQCs becomes simple. While the simpler PQC typically means fewer quantum operations and accumulated control errors, thus corresponding to a smaller $\text{Var}(o_k)$. So we can deal with the tradeoff according to practical experimental conditions to reach the optimal precision.

Besides, though the minimum number of readout operations required to reconstruct STR quantum states is given by Eq. \eqref{minN}, additional ones can be introduced to improve the precision. The average and the standard deviation of $f(\Theta)$ over 100 sets of random PQCs are shown in Fig.\ \ref{othStr}(b), where the $x$-axis labels the additional PQCs in each set. Consequently, additional experimental measurements can be employed when their time consumption is acceptable.

\section{Conclusion}

We demonstrated that the full quantum state tomography of star-topology register is feasible even though the controllability and measurement of peripheral spins are constrained. Utilizing the star-symmetry of STR, we design a compact strategy with the minimum number of measurements, which scales polynomially with the size of system. We further quantify the precision of tomography caused in noisy experimental measurements. By optimizing the PQCs for transferring information, the robustness against noise can be improved. The presented 10-spin example confirmed the feasibility and scalability of our scheme. In this case, STR quantum state can be fully determined with no more than 37 readout operations and more than three quarters decrease of total variance. 

Our approach is a promising tool for the tomography of other quantum system with constrained controllability and measurements. Besides, due to the readout operations generated by PQCs are flexible to be adjusted, they can be further optimized according to the specific experimental conditions, such as measurement noise \cite{meaErr}, control errors \cite{controlerr3,controlerr4} and relaxation effect \cite{decErr}. Our work also has potential applications in multiparameter quantum metrology \cite{multipara}, quantum enhanced imaging \cite{imag1,imag2} and various quantum-process-tomography experiments \cite{protomo1,protomo2,protomo3,zeroNMR2}.

\begin{acknowledgements}
This work is supported by the National Key R \& D
Program of China (Grants No. 2018YFA0306600 and
2016YFA0301203), the National Science Foundation of China
(Grants No. 11822502, 11974125 and 11927811), Anhui
Initiative in Quantum Information Technologies (Grant No.
AHY050000), the National Natural Science Foundation of China (Grants No. 92165108), USTC Research Funds of the Double First-Class Initiative and Anhui Provincial Natural Science Foundation (2108085J04). 
\end{acknowledgements}
	\bibliographystyle{apsrev4-2}
	\bibliography{Reference}

\providecommand{\noopsort}[1]{}\providecommand{\singleletter}[1]{#1}%
\begin{thebibliography}{48}%
\makeatletter
\providecommand \@ifxundefined [1]{%
 \@ifx{#1\undefined}
}%
\providecommand \@ifnum [1]{%
 \ifnum #1\expandafter \@firstoftwo
 \else \expandafter \@secondoftwo
 \fi
}%
\providecommand \@ifx [1]{%
 \ifx #1\expandafter \@firstoftwo
 \else \expandafter \@secondoftwo
 \fi
}%
\providecommand \natexlab [1]{#1}%
\providecommand \enquote  [1]{``#1''}%
\providecommand \bibnamefont  [1]{#1}%
\providecommand \bibfnamefont [1]{#1}%
\providecommand \citenamefont [1]{#1}%
\providecommand \href@noop [0]{\@secondoftwo}%
\providecommand \href [0]{\begingroup \@sanitize@url \@href}%
\providecommand \@href[1]{\@@startlink{#1}\@@href}%
\providecommand \@@href[1]{\endgroup#1\@@endlink}%
\providecommand \@sanitize@url [0]{\catcode `\\12\catcode `\$12\catcode
  `\&12\catcode `\#12\catcode `\^12\catcode `\_12\catcode `\%12\relax}%
\providecommand \@@startlink[1]{}%
\providecommand \@@endlink[0]{}%
\providecommand \url  [0]{\begingroup\@sanitize@url \@url }%
\providecommand \@url [1]{\endgroup\@href {#1}{\urlprefix }}%
\providecommand \urlprefix  [0]{URL }%
\providecommand \Eprint [0]{\href }%
\providecommand \doibase [0]{https://doi.org/}%
\providecommand \selectlanguage [0]{\@gobble}%
\providecommand \bibinfo  [0]{\@secondoftwo}%
\providecommand \bibfield  [0]{\@secondoftwo}%
\providecommand \translation [1]{[#1]}%
\providecommand \BibitemOpen [0]{}%
\providecommand \bibitemStop [0]{}%
\providecommand \bibitemNoStop [0]{.\EOS\space}%
\providecommand \EOS [0]{\spacefactor3000\relax}%
\providecommand \BibitemShut  [1]{\csname bibitem#1\endcsname}%
\let\auto@bib@innerbib\@empty
\bibitem [{\citenamefont {Paris}\ and\ \citenamefont {Rehacek}(2004)}]{BK1}%
  \BibitemOpen
  \bibfield  {author} {\bibinfo {author} {\bibfnamefont {M.}~\bibnamefont
  {Paris}}\ and\ \bibinfo {author} {\bibfnamefont {J.}~\bibnamefont
  {Rehacek}},\ }\href@noop {} {\emph {\bibinfo {title} {Quantum state
  estimation}}},\ Vol.\ \bibinfo {volume} {649}\ (\bibinfo  {publisher}
  {Springer Science \& Business Media},\ \bibinfo {year} {2004})\BibitemShut
  {NoStop}%
\bibitem [{\citenamefont {Welsch}\ \emph {et~al.}(1999)\citenamefont {Welsch},
  \citenamefont {Vogel},\ and\ \citenamefont {Opatrný}}]{BK2}%
  \BibitemOpen
  \bibfield  {author} {\bibinfo {author} {\bibfnamefont {D.-G.}\ \bibnamefont
  {Welsch}}, \bibinfo {author} {\bibfnamefont {W.}~\bibnamefont {Vogel}},\ and\
  \bibinfo {author} {\bibfnamefont {T.}~\bibnamefont {Opatrný}},\ }\bibinfo
  {title} {Ii homodyne detection and quantum-state reconstruction},\ in\ \href
  {https://doi.org/10.1016/s0079-6638(08)70389-5} {\emph {\bibinfo {booktitle}
  {Progress in Optics}}},\ \bibinfo {series} {Progress in Optics},
  Vol.~\bibinfo {volume} {39},\ \bibinfo {editor} {edited by\ \bibinfo {editor}
  {\bibfnamefont {E.}~\bibnamefont {Wolf}}}\ (\bibinfo  {publisher}
  {Elsevier},\ \bibinfo {year} {1999})\ pp.\ \bibinfo {pages}
  {63--211}\BibitemShut {NoStop}%
\bibitem [{\citenamefont {Leonhardt}(1997)}]{BK3}%
  \BibitemOpen
  \bibfield  {author} {\bibinfo {author} {\bibfnamefont {U.}~\bibnamefont
  {Leonhardt}},\ }\href@noop {} {\emph {\bibinfo {title} {Measuring the quantum
  state of light}}},\ Vol.~\bibinfo {volume} {22}\ (\bibinfo  {publisher}
  {Cambridge university press},\ \bibinfo {year} {1997})\BibitemShut {NoStop}%
\bibitem [{\citenamefont {Lu}\ \emph {et~al.}(2016)\citenamefont {Lu},
  \citenamefont {Xin}, \citenamefont {Yu}, \citenamefont {Ji}, \citenamefont
  {Chen}, \citenamefont {Long}, \citenamefont {Baugh}, \citenamefont {Peng},
  \citenamefont {Zeng},\ and\ \citenamefont {Laflamme}}]{BK4}%
  \BibitemOpen
  \bibfield  {author} {\bibinfo {author} {\bibfnamefont {D.}~\bibnamefont
  {Lu}}, \bibinfo {author} {\bibfnamefont {T.}~\bibnamefont {Xin}}, \bibinfo
  {author} {\bibfnamefont {N.}~\bibnamefont {Yu}}, \bibinfo {author}
  {\bibfnamefont {Z.}~\bibnamefont {Ji}}, \bibinfo {author} {\bibfnamefont
  {J.}~\bibnamefont {Chen}}, \bibinfo {author} {\bibfnamefont {G.}~\bibnamefont
  {Long}}, \bibinfo {author} {\bibfnamefont {J.}~\bibnamefont {Baugh}},
  \bibinfo {author} {\bibfnamefont {X.}~\bibnamefont {Peng}}, \bibinfo {author}
  {\bibfnamefont {B.}~\bibnamefont {Zeng}},\ and\ \bibinfo {author}
  {\bibfnamefont {R.}~\bibnamefont {Laflamme}},\ }\href
  {https://doi.org/10.1103/PhysRevLett.116.230501} {\bibfield  {journal}
  {\bibinfo  {journal} {Phys Rev Lett}\ }\textbf {\bibinfo {volume} {116}},\
  \bibinfo {pages} {230501} (\bibinfo {year} {2016})}\BibitemShut {NoStop}%
\bibitem [{\citenamefont {Li}\ \emph {et~al.}(2017)\citenamefont {Li},
  \citenamefont {Huang}, \citenamefont {Luo}, \citenamefont {Li}, \citenamefont
  {Lu},\ and\ \citenamefont {Zeng}}]{Litomo}%
  \BibitemOpen
  \bibfield  {author} {\bibinfo {author} {\bibfnamefont {J.}~\bibnamefont
  {Li}}, \bibinfo {author} {\bibfnamefont {S.}~\bibnamefont {Huang}}, \bibinfo
  {author} {\bibfnamefont {Z.}~\bibnamefont {Luo}}, \bibinfo {author}
  {\bibfnamefont {K.}~\bibnamefont {Li}}, \bibinfo {author} {\bibfnamefont
  {D.}~\bibnamefont {Lu}},\ and\ \bibinfo {author} {\bibfnamefont
  {B.}~\bibnamefont {Zeng}},\ }\href
  {https://doi.org/10.1103/PhysRevA.96.032307} {\bibfield  {journal} {\bibinfo
  {journal} {Physical Review A}\ }\textbf {\bibinfo {volume} {96}},\ \bibinfo
  {pages} {032307} (\bibinfo {year} {2017})}\BibitemShut {NoStop}%
\bibitem [{\citenamefont {Yang}\ \emph {et~al.}(2020)\citenamefont {Yang},
  \citenamefont {Yu}, \citenamefont {Betzholz}, \citenamefont {Arenz},\ and\
  \citenamefont {Cai}}]{fullcontrl}%
  \BibitemOpen
  \bibfield  {author} {\bibinfo {author} {\bibfnamefont {P.}~\bibnamefont
  {Yang}}, \bibinfo {author} {\bibfnamefont {M.}~\bibnamefont {Yu}}, \bibinfo
  {author} {\bibfnamefont {R.}~\bibnamefont {Betzholz}}, \bibinfo {author}
  {\bibfnamefont {C.}~\bibnamefont {Arenz}},\ and\ \bibinfo {author}
  {\bibfnamefont {J.}~\bibnamefont {Cai}},\ }\href
  {https://doi.org/10.1103/PhysRevLett.124.010405} {\bibfield  {journal}
  {\bibinfo  {journal} {Phys Rev Lett}\ }\textbf {\bibinfo {volume} {124}},\
  \bibinfo {pages} {010405} (\bibinfo {year} {2020})}\BibitemShut {NoStop}%
\bibitem [{\citenamefont {Foletti}\ \emph {et~al.}(2009)\citenamefont
  {Foletti}, \citenamefont {Bluhm}, \citenamefont {Mahalu}, \citenamefont
  {Umansky},\ and\ \citenamefont {Yacoby}}]{fullcontrl2}%
  \BibitemOpen
  \bibfield  {author} {\bibinfo {author} {\bibfnamefont {S.}~\bibnamefont
  {Foletti}}, \bibinfo {author} {\bibfnamefont {H.}~\bibnamefont {Bluhm}},
  \bibinfo {author} {\bibfnamefont {D.}~\bibnamefont {Mahalu}}, \bibinfo
  {author} {\bibfnamefont {V.}~\bibnamefont {Umansky}},\ and\ \bibinfo {author}
  {\bibfnamefont {A.}~\bibnamefont {Yacoby}},\ }\href
  {https://doi.org/10.1038/Nphys1424} {\bibfield  {journal} {\bibinfo
  {journal} {Nature Physics}\ }\textbf {\bibinfo {volume} {5}},\ \bibinfo
  {pages} {903} (\bibinfo {year} {2009})}\BibitemShut {NoStop}%
\bibitem [{\citenamefont {Barreiro}\ \emph {et~al.}(2011)\citenamefont
  {Barreiro}, \citenamefont {Muller}, \citenamefont {Schindler}, \citenamefont
  {Nigg}, \citenamefont {Monz}, \citenamefont {Chwalla}, \citenamefont
  {Hennrich}, \citenamefont {Roos}, \citenamefont {Zoller},\ and\ \citenamefont
  {Blatt}}]{fullcontrl3}%
  \BibitemOpen
  \bibfield  {author} {\bibinfo {author} {\bibfnamefont {J.~T.}\ \bibnamefont
  {Barreiro}}, \bibinfo {author} {\bibfnamefont {M.}~\bibnamefont {Muller}},
  \bibinfo {author} {\bibfnamefont {P.}~\bibnamefont {Schindler}}, \bibinfo
  {author} {\bibfnamefont {D.}~\bibnamefont {Nigg}}, \bibinfo {author}
  {\bibfnamefont {T.}~\bibnamefont {Monz}}, \bibinfo {author} {\bibfnamefont
  {M.}~\bibnamefont {Chwalla}}, \bibinfo {author} {\bibfnamefont
  {M.}~\bibnamefont {Hennrich}}, \bibinfo {author} {\bibfnamefont {C.~F.}\
  \bibnamefont {Roos}}, \bibinfo {author} {\bibfnamefont {P.}~\bibnamefont
  {Zoller}},\ and\ \bibinfo {author} {\bibfnamefont {R.}~\bibnamefont
  {Blatt}},\ }\href {https://doi.org/10.1038/nature09801} {\bibfield  {journal}
  {\bibinfo  {journal} {Nature}\ }\textbf {\bibinfo {volume} {470}},\ \bibinfo
  {pages} {486} (\bibinfo {year} {2011})}\BibitemShut {NoStop}%
\bibitem [{\citenamefont {Hong}\ \emph {et~al.}(1987)\citenamefont {Hong},
  \citenamefont {Ou},\ and\ \citenamefont {Mandel}}]{Hongou}%
  \BibitemOpen
  \bibfield  {author} {\bibinfo {author} {\bibfnamefont {C.~K.}\ \bibnamefont
  {Hong}}, \bibinfo {author} {\bibfnamefont {Z.~Y.}\ \bibnamefont {Ou}},\ and\
  \bibinfo {author} {\bibfnamefont {L.}~\bibnamefont {Mandel}},\ }\href
  {https://doi.org/10.1103/PhysRevLett.59.2044} {\bibfield  {journal} {\bibinfo
   {journal} {Phys Rev Lett}\ }\textbf {\bibinfo {volume} {59}},\ \bibinfo
  {pages} {2044} (\bibinfo {year} {1987})}\BibitemShut {NoStop}%
\bibitem [{\citenamefont {Bohnet}\ \emph {et~al.}(2016)\citenamefont {Bohnet},
  \citenamefont {Sawyer}, \citenamefont {Britton}, \citenamefont {Wall},
  \citenamefont {Rey}, \citenamefont {Foss-Feig},\ and\ \citenamefont
  {Bollinger}}]{trapped1}%
  \BibitemOpen
  \bibfield  {author} {\bibinfo {author} {\bibfnamefont {J.~G.}\ \bibnamefont
  {Bohnet}}, \bibinfo {author} {\bibfnamefont {B.~C.}\ \bibnamefont {Sawyer}},
  \bibinfo {author} {\bibfnamefont {J.~W.}\ \bibnamefont {Britton}}, \bibinfo
  {author} {\bibfnamefont {M.~L.}\ \bibnamefont {Wall}}, \bibinfo {author}
  {\bibfnamefont {A.~M.}\ \bibnamefont {Rey}}, \bibinfo {author} {\bibfnamefont
  {M.}~\bibnamefont {Foss-Feig}},\ and\ \bibinfo {author} {\bibfnamefont
  {J.~J.}\ \bibnamefont {Bollinger}},\ }\href
  {https://doi.org/10.1126/science.aad9958} {\bibfield  {journal} {\bibinfo
  {journal} {Science}\ }\textbf {\bibinfo {volume} {352}},\ \bibinfo {pages}
  {1297} (\bibinfo {year} {2016})}\BibitemShut {NoStop}%
\bibitem [{\citenamefont {Britton}\ \emph {et~al.}(2012)\citenamefont
  {Britton}, \citenamefont {Sawyer}, \citenamefont {Keith}, \citenamefont
  {Wang}, \citenamefont {Freericks}, \citenamefont {Uys}, \citenamefont
  {Biercuk},\ and\ \citenamefont {Bollinger}}]{trapped2}%
  \BibitemOpen
  \bibfield  {author} {\bibinfo {author} {\bibfnamefont {J.~W.}\ \bibnamefont
  {Britton}}, \bibinfo {author} {\bibfnamefont {B.~C.}\ \bibnamefont {Sawyer}},
  \bibinfo {author} {\bibfnamefont {A.~C.}\ \bibnamefont {Keith}}, \bibinfo
  {author} {\bibfnamefont {C.~C.}\ \bibnamefont {Wang}}, \bibinfo {author}
  {\bibfnamefont {J.~K.}\ \bibnamefont {Freericks}}, \bibinfo {author}
  {\bibfnamefont {H.}~\bibnamefont {Uys}}, \bibinfo {author} {\bibfnamefont
  {M.~J.}\ \bibnamefont {Biercuk}},\ and\ \bibinfo {author} {\bibfnamefont
  {J.~J.}\ \bibnamefont {Bollinger}},\ }\href
  {https://doi.org/10.1038/nature10981} {\bibfield  {journal} {\bibinfo
  {journal} {Nature}\ }\textbf {\bibinfo {volume} {484}},\ \bibinfo {pages}
  {489} (\bibinfo {year} {2012})}\BibitemShut {NoStop}%
\bibitem [{\citenamefont {Szymanski}(1988)}]{mageq1}%
  \BibitemOpen
  \bibfield  {author} {\bibinfo {author} {\bibfnamefont {S.}~\bibnamefont
  {Szymanski}},\ }\href {https://doi.org/10.1016/0022-2364(88)90181-3}
  {\bibfield  {journal} {\bibinfo  {journal} {Journal of Magnetic Resonance
  (1969)}\ }\textbf {\bibinfo {volume} {77}},\ \bibinfo {pages} {320} (\bibinfo
  {year} {1988})}\BibitemShut {NoStop}%
\bibitem [{\citenamefont {Abragam}(1961)}]{mageq2}%
  \BibitemOpen
  \bibfield  {author} {\bibinfo {author} {\bibfnamefont {A.}~\bibnamefont
  {Abragam}},\ }\href@noop {} {\emph {\bibinfo {title} {The principles of
  nuclear magnetism}}}\ (\bibinfo  {publisher} {Oxford university press},\
  \bibinfo {year} {1961})\BibitemShut {NoStop}%
\bibitem [{\citenamefont {Toth}\ \emph {et~al.}(2010)\citenamefont {Toth},
  \citenamefont {Wieczorek}, \citenamefont {Gross}, \citenamefont {Krischek},
  \citenamefont {Schwemmer},\ and\ \citenamefont {Weinfurter}}]{PI}%
  \BibitemOpen
  \bibfield  {author} {\bibinfo {author} {\bibfnamefont {G.}~\bibnamefont
  {Toth}}, \bibinfo {author} {\bibfnamefont {W.}~\bibnamefont {Wieczorek}},
  \bibinfo {author} {\bibfnamefont {D.}~\bibnamefont {Gross}}, \bibinfo
  {author} {\bibfnamefont {R.}~\bibnamefont {Krischek}}, \bibinfo {author}
  {\bibfnamefont {C.}~\bibnamefont {Schwemmer}},\ and\ \bibinfo {author}
  {\bibfnamefont {H.}~\bibnamefont {Weinfurter}},\ }\href
  {https://doi.org/10.1103/PhysRevLett.105.250403} {\bibfield  {journal}
  {\bibinfo  {journal} {Phys Rev Lett}\ }\textbf {\bibinfo {volume} {105}},\
  \bibinfo {pages} {250403} (\bibinfo {year} {2010})}\BibitemShut {NoStop}%
\bibitem [{\citenamefont {Karassiov}(2005)}]{PItomo1}%
  \BibitemOpen
  \bibfield  {author} {\bibinfo {author} {\bibfnamefont {V.~P.}\ \bibnamefont
  {Karassiov}},\ }\href {https://doi.org/DOI 10.1007/s10946-005-0047-8}
  {\bibfield  {journal} {\bibinfo  {journal} {Journal of Russian Laser
  Research}\ }\textbf {\bibinfo {volume} {26}},\ \bibinfo {pages} {484}
  (\bibinfo {year} {2005})}\BibitemShut {NoStop}%
\bibitem [{\citenamefont {Adamson}\ \emph {et~al.}(2007)\citenamefont
  {Adamson}, \citenamefont {Shalm}, \citenamefont {Mitchell},\ and\
  \citenamefont {Steinberg}}]{hidden}%
  \BibitemOpen
  \bibfield  {author} {\bibinfo {author} {\bibfnamefont {R.~B.}\ \bibnamefont
  {Adamson}}, \bibinfo {author} {\bibfnamefont {L.~K.}\ \bibnamefont {Shalm}},
  \bibinfo {author} {\bibfnamefont {M.~W.}\ \bibnamefont {Mitchell}},\ and\
  \bibinfo {author} {\bibfnamefont {A.~M.}\ \bibnamefont {Steinberg}},\ }\href
  {https://doi.org/10.1103/PhysRevLett.98.043601} {\bibfield  {journal}
  {\bibinfo  {journal} {Phys Rev Lett}\ }\textbf {\bibinfo {volume} {98}},\
  \bibinfo {pages} {043601} (\bibinfo {year} {2007})}\BibitemShut {NoStop}%
\bibitem [{\citenamefont {Adamson}\ \emph {et~al.}(2008)\citenamefont
  {Adamson}, \citenamefont {Turner}, \citenamefont {Mitchell},\ and\
  \citenamefont {Steinberg}}]{PItomo2}%
  \BibitemOpen
  \bibfield  {author} {\bibinfo {author} {\bibfnamefont {R.~B.~A.}\
  \bibnamefont {Adamson}}, \bibinfo {author} {\bibfnamefont {P.~S.}\
  \bibnamefont {Turner}}, \bibinfo {author} {\bibfnamefont {M.~W.}\
  \bibnamefont {Mitchell}},\ and\ \bibinfo {author} {\bibfnamefont {A.~M.}\
  \bibnamefont {Steinberg}},\ }\href
  {https://doi.org/10.1103/PhysRevA.78.033832} {\bibfield  {journal} {\bibinfo
  {journal} {Physical Review A}\ }\textbf {\bibinfo {volume} {78}},\ \bibinfo
  {pages} {033832} (\bibinfo {year} {2008})}\BibitemShut {NoStop}%
\bibitem [{\citenamefont {Moroder}\ \emph {et~al.}(2012)\citenamefont
  {Moroder}, \citenamefont {Hyllus}, \citenamefont {Toth}, \citenamefont
  {Schwemmer}, \citenamefont {Niggebaum}, \citenamefont {Gaile}, \citenamefont
  {Guhne},\ and\ \citenamefont {Weinfurter}}]{PItomo3}%
  \BibitemOpen
  \bibfield  {author} {\bibinfo {author} {\bibfnamefont {T.}~\bibnamefont
  {Moroder}}, \bibinfo {author} {\bibfnamefont {P.}~\bibnamefont {Hyllus}},
  \bibinfo {author} {\bibfnamefont {G.}~\bibnamefont {Toth}}, \bibinfo {author}
  {\bibfnamefont {C.}~\bibnamefont {Schwemmer}}, \bibinfo {author}
  {\bibfnamefont {A.}~\bibnamefont {Niggebaum}}, \bibinfo {author}
  {\bibfnamefont {S.}~\bibnamefont {Gaile}}, \bibinfo {author} {\bibfnamefont
  {O.}~\bibnamefont {Guhne}},\ and\ \bibinfo {author} {\bibfnamefont
  {H.}~\bibnamefont {Weinfurter}},\ }\href {https://doi.org/Artn 105001
  10.1088/1367-2630/14/10/105001} {\bibfield  {journal} {\bibinfo  {journal}
  {New Journal of Physics}\ }\textbf {\bibinfo {volume} {14}},\ \bibinfo
  {pages} {105001} (\bibinfo {year} {2012})}\BibitemShut {NoStop}%
\bibitem [{\citenamefont {Banchi}\ \emph {et~al.}(2018)\citenamefont {Banchi},
  \citenamefont {Kolthammer},\ and\ \citenamefont {Kim}}]{PItomo5}%
  \BibitemOpen
  \bibfield  {author} {\bibinfo {author} {\bibfnamefont {L.}~\bibnamefont
  {Banchi}}, \bibinfo {author} {\bibfnamefont {W.~S.}\ \bibnamefont
  {Kolthammer}},\ and\ \bibinfo {author} {\bibfnamefont {M.~S.}\ \bibnamefont
  {Kim}},\ }\href {https://doi.org/10.1103/PhysRevLett.121.250402} {\bibfield
  {journal} {\bibinfo  {journal} {Phys Rev Lett}\ }\textbf {\bibinfo {volume}
  {121}},\ \bibinfo {pages} {250402} (\bibinfo {year} {2018})}\BibitemShut
  {NoStop}%
\bibitem [{\citenamefont {Jones}\ \emph {et~al.}(2009)\citenamefont {Jones},
  \citenamefont {Karlen}, \citenamefont {Fitzsimons}, \citenamefont {Ardavan},
  \citenamefont {Benjamin}, \citenamefont {Briggs},\ and\ \citenamefont
  {Morton}}]{apy1}%
  \BibitemOpen
  \bibfield  {author} {\bibinfo {author} {\bibfnamefont {J.~A.}\ \bibnamefont
  {Jones}}, \bibinfo {author} {\bibfnamefont {S.~D.}\ \bibnamefont {Karlen}},
  \bibinfo {author} {\bibfnamefont {J.}~\bibnamefont {Fitzsimons}}, \bibinfo
  {author} {\bibfnamefont {A.}~\bibnamefont {Ardavan}}, \bibinfo {author}
  {\bibfnamefont {S.~C.}\ \bibnamefont {Benjamin}}, \bibinfo {author}
  {\bibfnamefont {G.~A.}\ \bibnamefont {Briggs}},\ and\ \bibinfo {author}
  {\bibfnamefont {J.~J.}\ \bibnamefont {Morton}},\ }\href
  {https://doi.org/10.1126/science.1170730} {\bibfield  {journal} {\bibinfo
  {journal} {Science}\ }\textbf {\bibinfo {volume} {324}},\ \bibinfo {pages}
  {1166} (\bibinfo {year} {2009})}\BibitemShut {NoStop}%
\bibitem [{\citenamefont {Shukla}\ \emph {et~al.}(2014)\citenamefont {Shukla},
  \citenamefont {Sharma},\ and\ \citenamefont {Mahesh}}]{apy2}%
  \BibitemOpen
  \bibfield  {author} {\bibinfo {author} {\bibfnamefont {A.}~\bibnamefont
  {Shukla}}, \bibinfo {author} {\bibfnamefont {M.}~\bibnamefont {Sharma}},\
  and\ \bibinfo {author} {\bibfnamefont {T.~S.}\ \bibnamefont {Mahesh}},\
  }\href {https://doi.org/10.1016/j.cplett.2013.11.065} {\bibfield  {journal}
  {\bibinfo  {journal} {Chemical Physics Letters}\ }\textbf {\bibinfo {volume}
  {592}},\ \bibinfo {pages} {227} (\bibinfo {year} {2014})}\BibitemShut
  {NoStop}%
\bibitem [{\citenamefont {Khurana}\ \emph {et~al.}(2016)\citenamefont
  {Khurana}, \citenamefont {Unnikrishnan},\ and\ \citenamefont
  {Mahesh}}]{apy3}%
  \BibitemOpen
  \bibfield  {author} {\bibinfo {author} {\bibfnamefont {D.}~\bibnamefont
  {Khurana}}, \bibinfo {author} {\bibfnamefont {G.}~\bibnamefont
  {Unnikrishnan}},\ and\ \bibinfo {author} {\bibfnamefont {T.~S.}\ \bibnamefont
  {Mahesh}},\ }\href {https://doi.org/10.1103/PhysRevA.94.062334} {\bibfield
  {journal} {\bibinfo  {journal} {Physical Review A}\ }\textbf {\bibinfo
  {volume} {94}},\ \bibinfo {pages} {062334} (\bibinfo {year}
  {2016})}\BibitemShut {NoStop}%
\bibitem [{\citenamefont {Peng}\ \emph {et~al.}(2015)\citenamefont {Peng},
  \citenamefont {Zhou}, \citenamefont {Wei}, \citenamefont {Cui}, \citenamefont
  {Du},\ and\ \citenamefont {Liu}}]{apy4}%
  \BibitemOpen
  \bibfield  {author} {\bibinfo {author} {\bibfnamefont {X.}~\bibnamefont
  {Peng}}, \bibinfo {author} {\bibfnamefont {H.}~\bibnamefont {Zhou}}, \bibinfo
  {author} {\bibfnamefont {B.~B.}\ \bibnamefont {Wei}}, \bibinfo {author}
  {\bibfnamefont {J.}~\bibnamefont {Cui}}, \bibinfo {author} {\bibfnamefont
  {J.}~\bibnamefont {Du}},\ and\ \bibinfo {author} {\bibfnamefont {R.~B.}\
  \bibnamefont {Liu}},\ }\href {https://doi.org/10.1103/PhysRevLett.114.010601}
  {\bibfield  {journal} {\bibinfo  {journal} {Phys Rev Lett}\ }\textbf
  {\bibinfo {volume} {114}},\ \bibinfo {pages} {010601} (\bibinfo {year}
  {2015})}\BibitemShut {NoStop}%
\bibitem [{\citenamefont {Pande}\ \emph {et~al.}(2017)\citenamefont {Pande},
  \citenamefont {Bhole}, \citenamefont {Khurana},\ and\ \citenamefont
  {Mahesh}}]{apy5}%
  \BibitemOpen
  \bibfield  {author} {\bibinfo {author} {\bibfnamefont {V.~R.}\ \bibnamefont
  {Pande}}, \bibinfo {author} {\bibfnamefont {G.}~\bibnamefont {Bhole}},
  \bibinfo {author} {\bibfnamefont {D.}~\bibnamefont {Khurana}},\ and\ \bibinfo
  {author} {\bibfnamefont {T.~S.}\ \bibnamefont {Mahesh}},\ }\href
  {https://doi.org/10.1103/PhysRevA.96.012330} {\bibfield  {journal} {\bibinfo
  {journal} {Physical Review A}\ }\textbf {\bibinfo {volume} {96}},\ \bibinfo
  {pages} {012330} (\bibinfo {year} {2017})}\BibitemShut {NoStop}%
\bibitem [{\citenamefont {Pal}\ \emph {et~al.}(2018)\citenamefont {Pal},
  \citenamefont {Nishad}, \citenamefont {Mahesh},\ and\ \citenamefont
  {Sreejith}}]{apy6}%
  \BibitemOpen
  \bibfield  {author} {\bibinfo {author} {\bibfnamefont {S.}~\bibnamefont
  {Pal}}, \bibinfo {author} {\bibfnamefont {N.}~\bibnamefont {Nishad}},
  \bibinfo {author} {\bibfnamefont {T.~S.}\ \bibnamefont {Mahesh}},\ and\
  \bibinfo {author} {\bibfnamefont {G.~J.}\ \bibnamefont {Sreejith}},\ }\href
  {https://doi.org/10.1103/PhysRevLett.120.180602} {\bibfield  {journal}
  {\bibinfo  {journal} {Phys Rev Lett}\ }\textbf {\bibinfo {volume} {120}},\
  \bibinfo {pages} {180602} (\bibinfo {year} {2018})}\BibitemShut {NoStop}%
\bibitem [{\citenamefont {Mahesh}\ \emph {et~al.}(2021)\citenamefont {Mahesh},
  \citenamefont {Khurana}, \citenamefont {Krithika}, \citenamefont {Sreejith},\
  and\ \citenamefont {Sudheer~Kumar}}]{sum}%
  \BibitemOpen
  \bibfield  {author} {\bibinfo {author} {\bibfnamefont {T.~S.}\ \bibnamefont
  {Mahesh}}, \bibinfo {author} {\bibfnamefont {D.}~\bibnamefont {Khurana}},
  \bibinfo {author} {\bibfnamefont {V.~R.}\ \bibnamefont {Krithika}}, \bibinfo
  {author} {\bibfnamefont {G.~J.}\ \bibnamefont {Sreejith}},\ and\ \bibinfo
  {author} {\bibfnamefont {C.~S.}\ \bibnamefont {Sudheer~Kumar}},\ }\bibfield
  {journal} {\bibinfo  {journal} {J Phys Condens Matter}\ }\textbf {\bibinfo
  {volume} {33}},\ \href {https://doi.org/10.1088/1361-648X/ac0dd3}
  {10.1088/1361-648X/ac0dd3} (\bibinfo {year} {2021})\BibitemShut {NoStop}%
\bibitem [{\citenamefont {Chen}\ \emph {et~al.}(2020)\citenamefont {Chen},
  \citenamefont {Zhou}, \citenamefont {Bian}, \citenamefont {Li},\ and\
  \citenamefont {Peng}}]{jiahui}%
  \BibitemOpen
  \bibfield  {author} {\bibinfo {author} {\bibfnamefont {J.}~\bibnamefont
  {Chen}}, \bibinfo {author} {\bibfnamefont {Y.}~\bibnamefont {Zhou}}, \bibinfo
  {author} {\bibfnamefont {J.}~\bibnamefont {Bian}}, \bibinfo {author}
  {\bibfnamefont {J.}~\bibnamefont {Li}},\ and\ \bibinfo {author}
  {\bibfnamefont {X.}~\bibnamefont {Peng}},\ }\href
  {https://doi.org/10.1103/PhysRevA.102.032602} {\bibfield  {journal} {\bibinfo
   {journal} {Physical Review A}\ }\textbf {\bibinfo {volume} {102}},\ \bibinfo
  {pages} {032602} (\bibinfo {year} {2020})}\BibitemShut {NoStop}%
\bibitem [{\citenamefont {Bacon}\ \emph {et~al.}(2006)\citenamefont {Bacon},
  \citenamefont {Chuang},\ and\ \citenamefont {Harrow}}]{Schur}%
  \BibitemOpen
  \bibfield  {author} {\bibinfo {author} {\bibfnamefont {D.}~\bibnamefont
  {Bacon}}, \bibinfo {author} {\bibfnamefont {I.~L.}\ \bibnamefont {Chuang}},\
  and\ \bibinfo {author} {\bibfnamefont {A.~W.}\ \bibnamefont {Harrow}},\
  }\href {https://doi.org/10.1103/PhysRevLett.97.170502} {\bibfield  {journal}
  {\bibinfo  {journal} {Phys Rev Lett}\ }\textbf {\bibinfo {volume} {97}},\
  \bibinfo {pages} {170502} (\bibinfo {year} {2006})}\BibitemShut {NoStop}%
\bibitem [{\citenamefont {Tolimieri}\ and\ \citenamefont
  {An}(1997)}]{distribu}%
  \BibitemOpen
  \bibfield  {author} {\bibinfo {author} {\bibfnamefont {R.}~\bibnamefont
  {Tolimieri}}\ and\ \bibinfo {author} {\bibfnamefont {M.}~\bibnamefont {An}},\
  }\href@noop {} {\emph {\bibinfo {title} {Time-frequency representations}}}\
  (\bibinfo  {publisher} {Springer Science \& Business Media},\ \bibinfo {year}
  {1997})\BibitemShut {NoStop}%
\bibitem [{\citenamefont {Lee}(2002)}]{NMRtomo}%
  \BibitemOpen
  \bibfield  {author} {\bibinfo {author} {\bibfnamefont {J.-S.}\ \bibnamefont
  {Lee}},\ }\href {https://doi.org/10.1016/s0375-9601(02)01479-2} {\bibfield
  {journal} {\bibinfo  {journal} {Physics Letters A}\ }\textbf {\bibinfo
  {volume} {305}},\ \bibinfo {pages} {349} (\bibinfo {year}
  {2002})}\BibitemShut {NoStop}%
\bibitem [{\citenamefont {Leskowitz}\ and\ \citenamefont
  {Mueller}(2004)}]{greed}%
  \BibitemOpen
  \bibfield  {author} {\bibinfo {author} {\bibfnamefont {G.~M.}\ \bibnamefont
  {Leskowitz}}\ and\ \bibinfo {author} {\bibfnamefont {L.~J.}\ \bibnamefont
  {Mueller}},\ }\href {https://doi.org/ARTN 052302 10.1103/PhysRevA.69.052302}
  {\bibfield  {journal} {\bibinfo  {journal} {Physical Review A}\ }\textbf
  {\bibinfo {volume} {69}},\ \bibinfo {pages} {052302} (\bibinfo {year}
  {2004})}\BibitemShut {NoStop}%
\bibitem [{\citenamefont {Ohliger}\ \emph {et~al.}(2013)\citenamefont
  {Ohliger}, \citenamefont {Nesme},\ and\ \citenamefont {Eisert}}]{PQC0}%
  \BibitemOpen
  \bibfield  {author} {\bibinfo {author} {\bibfnamefont {M.}~\bibnamefont
  {Ohliger}}, \bibinfo {author} {\bibfnamefont {V.}~\bibnamefont {Nesme}},\
  and\ \bibinfo {author} {\bibfnamefont {J.}~\bibnamefont {Eisert}},\ }\href
  {https://doi.org/Artn 015024 10.1088/1367-2630/15/1/015024} {\bibfield
  {journal} {\bibinfo  {journal} {New Journal of Physics}\ }\textbf {\bibinfo
  {volume} {15}},\ \bibinfo {pages} {015024} (\bibinfo {year}
  {2013})}\BibitemShut {NoStop}%
\bibitem [{\citenamefont {Schwemmer}\ \emph {et~al.}(2015)\citenamefont
  {Schwemmer}, \citenamefont {Knips}, \citenamefont {Richart}, \citenamefont
  {Weinfurter}, \citenamefont {Moroder}, \citenamefont {Kleinmann},\ and\
  \citenamefont {Guhne}}]{LI}%
  \BibitemOpen
  \bibfield  {author} {\bibinfo {author} {\bibfnamefont {C.}~\bibnamefont
  {Schwemmer}}, \bibinfo {author} {\bibfnamefont {L.}~\bibnamefont {Knips}},
  \bibinfo {author} {\bibfnamefont {D.}~\bibnamefont {Richart}}, \bibinfo
  {author} {\bibfnamefont {H.}~\bibnamefont {Weinfurter}}, \bibinfo {author}
  {\bibfnamefont {T.}~\bibnamefont {Moroder}}, \bibinfo {author} {\bibfnamefont
  {M.}~\bibnamefont {Kleinmann}},\ and\ \bibinfo {author} {\bibfnamefont
  {O.}~\bibnamefont {Guhne}},\ }\href
  {https://doi.org/10.1103/PhysRevLett.114.080403} {\bibfield  {journal}
  {\bibinfo  {journal} {Phys Rev Lett}\ }\textbf {\bibinfo {volume} {114}},\
  \bibinfo {pages} {080403} (\bibinfo {year} {2015})}\BibitemShut {NoStop}%
\bibitem [{\citenamefont {Moore}(1920)}]{Moore}%
  \BibitemOpen
  \bibfield  {author} {\bibinfo {author} {\bibfnamefont {E.~H.}\ \bibnamefont
  {Moore}},\ }\href@noop {} {\bibfield  {journal} {\bibinfo  {journal} {Bull.
  Am. Math. Soc.}\ }\textbf {\bibinfo {volume} {26}},\ \bibinfo {pages} {394}
  (\bibinfo {year} {1920})}\BibitemShut {NoStop}%
\bibitem [{\citenamefont {Nocedal}\ and\ \citenamefont {Wright}(2006)}]{SQP}%
  \BibitemOpen
  \bibfield  {author} {\bibinfo {author} {\bibfnamefont {J.}~\bibnamefont
  {Nocedal}}\ and\ \bibinfo {author} {\bibfnamefont {S.~J.}\ \bibnamefont
  {Wright}},\ }\href@noop {} {\bibfield  {journal} {\bibinfo  {journal}
  {Numerical optimization}\ ,\ \bibinfo {pages} {529}} (\bibinfo {year}
  {2006})}\BibitemShut {NoStop}%
\bibitem [{\citenamefont {Modi}\ \emph {et~al.}(2011)\citenamefont {Modi},
  \citenamefont {Cable}, \citenamefont {Williamson},\ and\ \citenamefont
  {Vedral}}]{mixedstt}%
  \BibitemOpen
  \bibfield  {author} {\bibinfo {author} {\bibfnamefont {K.}~\bibnamefont
  {Modi}}, \bibinfo {author} {\bibfnamefont {H.}~\bibnamefont {Cable}},
  \bibinfo {author} {\bibfnamefont {M.}~\bibnamefont {Williamson}},\ and\
  \bibinfo {author} {\bibfnamefont {V.}~\bibnamefont {Vedral}},\ }\href
  {https://doi.org/ARTN 021022 10.1103/PhysRevX.1.021022} {\bibfield  {journal}
  {\bibinfo  {journal} {Physical Review X}\ }\textbf {\bibinfo {volume} {1}},\
  \bibinfo {pages} {021022} (\bibinfo {year} {2011})}\BibitemShut {NoStop}%
\bibitem [{\citenamefont {Pezzè}\ \emph {et~al.}(2018)\citenamefont {Pezzè},
  \citenamefont {Smerzi}, \citenamefont {Oberthaler}, \citenamefont {Schmied},\
  and\ \citenamefont {Treutlein}}]{purestt}%
  \BibitemOpen
  \bibfield  {author} {\bibinfo {author} {\bibfnamefont {L.}~\bibnamefont
  {Pezzè}}, \bibinfo {author} {\bibfnamefont {A.}~\bibnamefont {Smerzi}},
  \bibinfo {author} {\bibfnamefont {M.~K.}\ \bibnamefont {Oberthaler}},
  \bibinfo {author} {\bibfnamefont {R.}~\bibnamefont {Schmied}},\ and\ \bibinfo
  {author} {\bibfnamefont {P.}~\bibnamefont {Treutlein}},\ }\href
  {https://doi.org/10.1103/RevModPhys.90.035005} {\bibfield  {journal}
  {\bibinfo  {journal} {Reviews of Modern Physics}\ }\textbf {\bibinfo {volume}
  {90}},\ \bibinfo {pages} {035005} (\bibinfo {year} {2018})}\BibitemShut
  {NoStop}%
\bibitem [{\citenamefont {Feng}\ \emph {et~al.}(2018)\citenamefont {Feng},
  \citenamefont {Cho}, \citenamefont {Katiyar}, \citenamefont {Li},
  \citenamefont {Lu}, \citenamefont {Baugh},\ and\ \citenamefont
  {Laflamme}}]{meaErr}%
  \BibitemOpen
  \bibfield  {author} {\bibinfo {author} {\bibfnamefont {G.}~\bibnamefont
  {Feng}}, \bibinfo {author} {\bibfnamefont {F.~H.}\ \bibnamefont {Cho}},
  \bibinfo {author} {\bibfnamefont {H.}~\bibnamefont {Katiyar}}, \bibinfo
  {author} {\bibfnamefont {J.}~\bibnamefont {Li}}, \bibinfo {author}
  {\bibfnamefont {D.}~\bibnamefont {Lu}}, \bibinfo {author} {\bibfnamefont
  {J.}~\bibnamefont {Baugh}},\ and\ \bibinfo {author} {\bibfnamefont
  {R.}~\bibnamefont {Laflamme}},\ }\href
  {https://doi.org/10.1103/PhysRevA.98.052341} {\bibfield  {journal} {\bibinfo
  {journal} {Physical Review A}\ }\textbf {\bibinfo {volume} {98}},\ \bibinfo
  {pages} {052341} (\bibinfo {year} {2018})}\BibitemShut {NoStop}%
\bibitem [{\citenamefont {Daems}\ \emph {et~al.}(2013)\citenamefont {Daems},
  \citenamefont {Ruschhaupt}, \citenamefont {Sugny},\ and\ \citenamefont
  {Guerin}}]{controlerr3}%
  \BibitemOpen
  \bibfield  {author} {\bibinfo {author} {\bibfnamefont {D.}~\bibnamefont
  {Daems}}, \bibinfo {author} {\bibfnamefont {A.}~\bibnamefont {Ruschhaupt}},
  \bibinfo {author} {\bibfnamefont {D.}~\bibnamefont {Sugny}},\ and\ \bibinfo
  {author} {\bibfnamefont {S.}~\bibnamefont {Guerin}},\ }\href
  {https://doi.org/10.1103/PhysRevLett.111.050404} {\bibfield  {journal}
  {\bibinfo  {journal} {Phys Rev Lett}\ }\textbf {\bibinfo {volume} {111}},\
  \bibinfo {pages} {050404} (\bibinfo {year} {2013})}\BibitemShut {NoStop}%
\bibitem [{\citenamefont {Souza}\ \emph {et~al.}(2011)\citenamefont {Souza},
  \citenamefont {Alvarez},\ and\ \citenamefont {Suter}}]{controlerr4}%
  \BibitemOpen
  \bibfield  {author} {\bibinfo {author} {\bibfnamefont {A.~M.}\ \bibnamefont
  {Souza}}, \bibinfo {author} {\bibfnamefont {G.~A.}\ \bibnamefont {Alvarez}},\
  and\ \bibinfo {author} {\bibfnamefont {D.}~\bibnamefont {Suter}},\ }\href
  {https://doi.org/10.1103/PhysRevLett.106.240501} {\bibfield  {journal}
  {\bibinfo  {journal} {Phys Rev Lett}\ }\textbf {\bibinfo {volume} {106}},\
  \bibinfo {pages} {240501} (\bibinfo {year} {2011})}\BibitemShut {NoStop}%
\bibitem [{\citenamefont {Negrevergne}\ \emph {et~al.}(2006)\citenamefont
  {Negrevergne}, \citenamefont {Mahesh}, \citenamefont {Ryan}, \citenamefont
  {Ditty}, \citenamefont {Cyr-Racine}, \citenamefont {Power}, \citenamefont
  {Boulant}, \citenamefont {Havel}, \citenamefont {Cory},\ and\ \citenamefont
  {Laflamme}}]{decErr}%
  \BibitemOpen
  \bibfield  {author} {\bibinfo {author} {\bibfnamefont {C.}~\bibnamefont
  {Negrevergne}}, \bibinfo {author} {\bibfnamefont {T.~S.}\ \bibnamefont
  {Mahesh}}, \bibinfo {author} {\bibfnamefont {C.~A.}\ \bibnamefont {Ryan}},
  \bibinfo {author} {\bibfnamefont {M.}~\bibnamefont {Ditty}}, \bibinfo
  {author} {\bibfnamefont {F.}~\bibnamefont {Cyr-Racine}}, \bibinfo {author}
  {\bibfnamefont {W.}~\bibnamefont {Power}}, \bibinfo {author} {\bibfnamefont
  {N.}~\bibnamefont {Boulant}}, \bibinfo {author} {\bibfnamefont
  {T.}~\bibnamefont {Havel}}, \bibinfo {author} {\bibfnamefont {D.~G.}\
  \bibnamefont {Cory}},\ and\ \bibinfo {author} {\bibfnamefont
  {R.}~\bibnamefont {Laflamme}},\ }\href
  {https://doi.org/10.1103/PhysRevLett.96.170501} {\bibfield  {journal}
  {\bibinfo  {journal} {Phys Rev Lett}\ }\textbf {\bibinfo {volume} {96}},\
  \bibinfo {pages} {170501} (\bibinfo {year} {2006})}\BibitemShut {NoStop}%
\bibitem [{\citenamefont {Meyer}\ \emph {et~al.}(2021)\citenamefont {Meyer},
  \citenamefont {Borregaard},\ and\ \citenamefont {Eisert}}]{multipara}%
  \BibitemOpen
  \bibfield  {author} {\bibinfo {author} {\bibfnamefont {J.~J.}\ \bibnamefont
  {Meyer}}, \bibinfo {author} {\bibfnamefont {J.}~\bibnamefont {Borregaard}},\
  and\ \bibinfo {author} {\bibfnamefont {J.}~\bibnamefont {Eisert}},\ }\href
  {https://doi.org/ARTN 89 10.1038/s41534-021-00425-y} {\bibfield  {journal}
  {\bibinfo  {journal} {Npj Quantum Information}\ }\textbf {\bibinfo {volume}
  {7}},\ \bibinfo {pages} {89} (\bibinfo {year} {2021})}\BibitemShut {NoStop}%
\bibitem [{\citenamefont {Perez-Delgado}\ \emph {et~al.}(2012)\citenamefont
  {Perez-Delgado}, \citenamefont {Pearce},\ and\ \citenamefont {Kok}}]{imag1}%
  \BibitemOpen
  \bibfield  {author} {\bibinfo {author} {\bibfnamefont {C.~A.}\ \bibnamefont
  {Perez-Delgado}}, \bibinfo {author} {\bibfnamefont {M.~E.}\ \bibnamefont
  {Pearce}},\ and\ \bibinfo {author} {\bibfnamefont {P.}~\bibnamefont {Kok}},\
  }\href {https://doi.org/10.1103/PhysRevLett.109.123601} {\bibfield  {journal}
  {\bibinfo  {journal} {Phys Rev Lett}\ }\textbf {\bibinfo {volume} {109}},\
  \bibinfo {pages} {123601} (\bibinfo {year} {2012})}\BibitemShut {NoStop}%
\bibitem [{\citenamefont {Humphreys}\ \emph {et~al.}(2013)\citenamefont
  {Humphreys}, \citenamefont {Barbieri}, \citenamefont {Datta},\ and\
  \citenamefont {Walmsley}}]{imag2}%
  \BibitemOpen
  \bibfield  {author} {\bibinfo {author} {\bibfnamefont {P.~C.}\ \bibnamefont
  {Humphreys}}, \bibinfo {author} {\bibfnamefont {M.}~\bibnamefont {Barbieri}},
  \bibinfo {author} {\bibfnamefont {A.}~\bibnamefont {Datta}},\ and\ \bibinfo
  {author} {\bibfnamefont {I.~A.}\ \bibnamefont {Walmsley}},\ }\href
  {https://doi.org/10.1103/PhysRevLett.111.070403} {\bibfield  {journal}
  {\bibinfo  {journal} {Phys Rev Lett}\ }\textbf {\bibinfo {volume} {111}},\
  \bibinfo {pages} {070403} (\bibinfo {year} {2013})}\BibitemShut {NoStop}%
\bibitem [{\citenamefont {Kim}\ \emph {et~al.}(2018)\citenamefont {Kim},
  \citenamefont {Kim}, \citenamefont {Lee}, \citenamefont {Han}, \citenamefont
  {Moon}, \citenamefont {Kim},\ and\ \citenamefont {Cho}}]{protomo1}%
  \BibitemOpen
  \bibfield  {author} {\bibinfo {author} {\bibfnamefont {Y.}~\bibnamefont
  {Kim}}, \bibinfo {author} {\bibfnamefont {Y.~S.}\ \bibnamefont {Kim}},
  \bibinfo {author} {\bibfnamefont {S.~Y.}\ \bibnamefont {Lee}}, \bibinfo
  {author} {\bibfnamefont {S.~W.}\ \bibnamefont {Han}}, \bibinfo {author}
  {\bibfnamefont {S.}~\bibnamefont {Moon}}, \bibinfo {author} {\bibfnamefont
  {Y.~H.}\ \bibnamefont {Kim}},\ and\ \bibinfo {author} {\bibfnamefont {Y.~W.}\
  \bibnamefont {Cho}},\ }\href {https://doi.org/10.1038/s41467-017-02511-2}
  {\bibfield  {journal} {\bibinfo  {journal} {Nat Commun}\ }\textbf {\bibinfo
  {volume} {9}},\ \bibinfo {pages} {192} (\bibinfo {year} {2018})}\BibitemShut
  {NoStop}%
\bibitem [{\citenamefont {Govia}\ \emph {et~al.}(2020)\citenamefont {Govia},
  \citenamefont {Ribeill}, \citenamefont {Riste}, \citenamefont {Ware},\ and\
  \citenamefont {Krovi}}]{protomo2}%
  \BibitemOpen
  \bibfield  {author} {\bibinfo {author} {\bibfnamefont {L.~C.~G.}\
  \bibnamefont {Govia}}, \bibinfo {author} {\bibfnamefont {G.~J.}\ \bibnamefont
  {Ribeill}}, \bibinfo {author} {\bibfnamefont {D.}~\bibnamefont {Riste}},
  \bibinfo {author} {\bibfnamefont {M.}~\bibnamefont {Ware}},\ and\ \bibinfo
  {author} {\bibfnamefont {H.}~\bibnamefont {Krovi}},\ }\href
  {https://doi.org/10.1038/s41467-020-14873-1} {\bibfield  {journal} {\bibinfo
  {journal} {Nat Commun}\ }\textbf {\bibinfo {volume} {11}},\ \bibinfo {pages}
  {1084} (\bibinfo {year} {2020})}\BibitemShut {NoStop}%
\bibitem [{\citenamefont {Maciel}\ \emph {et~al.}(2015)\citenamefont {Maciel},
  \citenamefont {Vianna}, \citenamefont {Sarthour},\ and\ \citenamefont
  {Oliveira}}]{protomo3}%
  \BibitemOpen
  \bibfield  {author} {\bibinfo {author} {\bibfnamefont {T.~O.}\ \bibnamefont
  {Maciel}}, \bibinfo {author} {\bibfnamefont {R.~O.}\ \bibnamefont {Vianna}},
  \bibinfo {author} {\bibfnamefont {R.~S.}\ \bibnamefont {Sarthour}},\ and\
  \bibinfo {author} {\bibfnamefont {I.~S.}\ \bibnamefont {Oliveira}},\ }\href
  {https://doi.org/10.1088/1367-2630/17/11/113012} {\bibfield  {journal}
  {\bibinfo  {journal} {New Journal of Physics}\ }\textbf {\bibinfo {volume}
  {17}},\ \bibinfo {pages} {113012} (\bibinfo {year} {2015})}\BibitemShut
  {NoStop}%
\bibitem [{\citenamefont {Jiang}\ \emph {et~al.}(2018)\citenamefont {Jiang},
  \citenamefont {Wu}, \citenamefont {Blanchard}, \citenamefont {Feng},
  \citenamefont {Peng},\ and\ \citenamefont {Budker}}]{zeroNMR2}%
  \BibitemOpen
  \bibfield  {author} {\bibinfo {author} {\bibfnamefont {M.}~\bibnamefont
  {Jiang}}, \bibinfo {author} {\bibfnamefont {T.}~\bibnamefont {Wu}}, \bibinfo
  {author} {\bibfnamefont {J.~W.}\ \bibnamefont {Blanchard}}, \bibinfo {author}
  {\bibfnamefont {G.}~\bibnamefont {Feng}}, \bibinfo {author} {\bibfnamefont
  {X.}~\bibnamefont {Peng}},\ and\ \bibinfo {author} {\bibfnamefont
  {D.}~\bibnamefont {Budker}},\ }\href {https://doi.org/10.1126/sciadv.aar6327}
  {\bibfield  {journal} {\bibinfo  {journal} {Sci Adv}\ }\textbf {\bibinfo
  {volume} {4}},\ \bibinfo {pages} {eaar6327} (\bibinfo {year}
  {2018})}\BibitemShut {NoStop}%
\end{thebibliography}%


\providecommand{\noopsort}[1]{}\providecommand{\singleletter}[1]{#1}%
%
\end{document}